\documentclass[aps,amsmath,amssymb,showkeys]{revtex4}
\usepackage[dvips]{graphicx,color}
\usepackage{times}
\usepackage{mathrsfs}
\usepackage{amsmath}
\usepackage{amssymb}
\usepackage{graphicx}
\usepackage{epsfig}
\usepackage{dcolumn}
\usepackage{bm}
\usepackage{xcolor}
\usepackage{hyperref}
\hypersetup{colorlinks=true, citecolor=blue, filecolor=red, urlcolor=cyan}
\begin{document}
\title{Galactic rotation dynamics in a new $f(\mathcal{R})$ gravity model}

\author{Nashiba Parbin}
\email[Email: ]{nashibaparbin91@gmail.com}
\author{Umananda Dev Goswami}
\email[Email: ]{umananda2@gmail.com}
\affiliation{Department of Physics, Dibrugarh University, Dibrugarh 786004, 
Assam, India}

\begin{abstract}
We propose to test the viability of the recently introduced $f(\mathcal{R})$ gravity  
model in the galactic scales. For this purpose we consider test particles 
moving in stable circular orbits around the galactic center. We study the 
Palatini approach of $f(\mathcal{R})$ gravity via Weyl transformation, which 
is the frame transformation from the Jordan frame to the Einstein frame. We 
derive the expression of rotational velocities of test particles in the 
new $f(\mathcal{R})$ gravity model. For the observational data of samples of high surface 
brightness and low surface brightness galaxies, we show that the predicted 
rotation curves are well fitted with observations, thus implying that this 
model can explain flat rotation curves of galaxies. We also study an ultra 
diffuse galaxy, AGC $242019$ which has been claimed in literature to be 
a dark matter dominated galaxy similar to low surface brightness galaxies 
with a slowly rising rotation curve. The rotation curve of this galaxy also 
fits well with the model prediction in our study. Furthermore, we studied the 
Tully-Fisher relation for the entire sample of galaxies and found that the 
model prediction shows the consistency with the data. 
\end{abstract}


\maketitle

\section{Introduction}
A longstanding challenge in astrophysics and cosmology is the mystery of dark 
matter (DM) \cite{bertone, swart, tim, frenk, strigari}. The missing mass 
problem was first predicted in the early 1930s by J.~H.~Oort \cite{oort} while 
studying the motion of stars in the Milky Way. Around the same time similar 
evidences were reported by Swiss astronomer Fritz Zwicky, who studied the Coma 
Cluster \cite{zwicky, Zwicky}. Two prime evidences that insist on the 
existence of DM are the velocity rotation curve of galaxies 
\cite{rubin, young, borriello} and the gravitational lensing 
\cite{young, massey}. American astronomer Vera Rubin, did pioneering work by 
conducting a study of rotation curves of 60 isolated galaxies \cite{garrett} 
and established the idea of ``missing mass''. Using the Planck data 
\cite{planck} on the Cosmic Microwave Background (CMB) radiation, 
measurements of the cosmological parameters infer that the Universe is made 
up of $\sim 4-5\%$ baryons, $\sim 25\%$ non-baryonic dark matter, and 
$\sim 70\%$ dark energy. A few particles claimed as DM candidates 
\cite{Bertone, feng} are, namely, weakly interacting massive particles 
(WIMPs), standard model (SM) neutrinos, sterile neutrinos, axions, 
supersymmetric candidates (neutralinos, sneutrinos, gravitinos, axinos), etc. 
However, after almost eight decades since the development of the concept of 
DM, the DM particle is still missing from the table of elementary particles of 
nature, i.e. the fundamental nature of DM remains a mystery, and the problem of DM persists.

Over the past few decades various other issues, especially the flatness and 
horizon problems \cite{guth}, and the riddle related to the late time cosmic 
acceleration \cite{reiss, perlmutter} have come to light, which specify 
that the standard cosmological model based on Einstein's General Relativity 
(GR) and the particle physics standard model fails to explain the Universe at 
large scales. This has generated an increasing interest to explore alternative 
theories of gravity (ATGs) \cite{Clifton, wagoner, ronald, yunes, alsing}, where gravitational interactions other 
than the ones described by GR were proposed. Within the broad area of ATGs, 
here we refer the modified theories of gravity (MTGs) \cite{capozziello, 
clifton, oikonomou, odintsov, sergei} as the ATGs those were proposed to 
modify GR to provide solutions to emerging issues. The simplest class of 
MTGs is the $f(\mathcal{R})$ gravity \cite{faraoni, felice}. In these gravity
theories the modification is made to the geometry part of Einstein's field 
equations. This is done by replacing the Ricci scalar $\mathcal{R}$ of the 
Einstein-Hilbert action with a function $f(\mathcal{R})$ of $\mathcal{R}$. There are two 
main variational approaches to derive the field equations in 
$f(\mathcal{R})$ gravity, the metric formalism and the palatini formalism. In 
the metric formalism, matter is minimally coupled with the metric, and the 
energy-momentum tensor is independently conserved. In Palatini formalism, the 
metric as well as the connection are treated as independent variables. Here, 
the Riemann tensor as well as the Ricci tensor are constructed with the 
independent connection. Moreover, there are other formalisms which are found
in the literature are the metric-affine formalism \cite{Sotiriou} and the
hybrid metric-Palatini formalism \cite{hybrid_metric_palatini}. In the 
metric-affine formalism the matter action is considered as variable with
respect to connection in contrast to the case of the Palatini formalism. The
hybrid metric-Palatini formalism is the combination of the suitable elements
of both metric and Palatini formalisms.   

A plethora of research works have been dedicated to explain the effects of DM 
in MTGs \cite{riazi, shtanov, nashiba, harko, yuri, katsuragawa, lobo}. 
As the concept of DM has mainly been indicated by irregularities in the 
galactic rotation curves, several MTGs have been proposed to study galactic 
rotation curves. Harko \cite{Harko} investigated galactic rotation 
curves in MTGs with non-minimal coupling between matter and geometry. 
Capozziello et al.~\cite{hybrid_metric_palatini} studied rotation curves in 
hybrid metric-Palatini gravity model. Gergely et al.~\cite{gergely} considered 
the asymptotic behaviour of galactic rotation curves in brane world models. 
Several other studies \cite{capone, sporea, finch, lin, troisi, salucci, 
martins, vipin} have been carried out in different MTGs to explain galactic 
rotation curves. These studies have motivated us to study and explain the 
rotation curves in a viable $f(\mathcal{R})$ gravity model. Here we propose to 
investigate the rotation curves of galaxies in the Palatini $f(\mathcal{R})$ 
gravity taking into consideration the conformal transformation of the 
metric. It is basically a frame transformation from the Jordan frame to the 
Einstein frame \cite{goswami, darabi, yuri_shtanov, olmo} via a conformal factor. To this aim, we consider the new model \cite{dhruba, Dhruba} of 
$f(\mathcal{R})$ gravity, which has been recently introduced as a viable dark
energy model of the theory. The main motive of the present study is to test 
the viability of this recent model in the galactic scales. We start by 
considering test particles around galaxies moving in stable circular orbits. 
The rotational velocity obtained by employing the new $f(\mathcal{R})$ gravity model has the 
Newtonian term as well as that coming from the modified geometry. The 
predicted rotational velocity from the model is fitted with observations of a 
few samples of high surface brightness (HSB), low surface brightness (LSB) and 
dwarf galaxies. Another addition to our work is the study of the ultra diffuse 
galaxies (UDGs). They are a fascinating class of galaxies with unusual 
properties, such as very high \cite{dokkum16}, or very low content of DM 
\cite{dokkum18}. Infact, UDGs are difficult to observe \cite{bothun91, 
bothun97, dokkum, roman} as well as to analyze \cite{ruiz, forbes}. We have 
investigated the behaviour of the rotation curve of a particular UDG, 
AGC $242019$ \cite{brook, shi21}. The rotational velocity predicted the model 
fits well for this UDG. Furthermore, we also derive the Tully-Fisher relation 
\cite{tully, brownstein, mcgaugh} for the new $f(\mathcal{R})$ gravity model.

Our work is organised as follows. In section \ref{sec.2}, we discuss the 
simplest type of modified gravity, i.e.~the $f(\mathcal{R})$ gravity in the 
Palatini formalism. Here, we obtain the modified field equations in 
$f(\mathcal{R})$ gravity. The Weyl transformation from the Jordan frame to 
the Einstein frame and its implications on the Palatini $f(\mathcal{R})$ 
gravity is also discussed in this section. In section \ref{sec.3}, we study 
the dynamics of a test particle around the centres of galaxies in conformally 
transformed Palatini $f(\mathcal{R})$ gravity. In section \ref{sec.4}, we 
derive the rotational velocity of the test particles in the recently proposed 
model of $f(\mathcal{R})$ gravity as mentioned above. 
In section \ref{sec.5}, we fit the predicted rotational velocity with 
observations as a test of the new $f(\mathcal{R})$ gravity model in galactic scales. Further, 
in section \ref{sec.6}, we derive the Tully-Fisher relation for the model. 
Finally, in section \ref{sec.7} we conclude and discuss the results of our 
work. Throughout our work we use the metric signature (- , +, +, +).

\section{Palatini $f(\mathcal{R})$ gravity and Weyl geometry}
\label{sec.2}

\subsection{Field Equations}

The action that defines $f(\mathcal{R})$ theories of gravity \cite{faraoni} 
has the generic form:
\begin{equation}
S = \frac{1}{2\kappa^2}\int d^4x \sqrt{-g}\,f(\mathcal{R}) + S_m\left[g^{\mu\nu},\Phi\right],
\label{eqn.1}
\end{equation}
where $f(\mathcal{R})$ is a function of the Ricci scalar $\mathcal{R} = g^{\mu\nu}\mathcal{R}_{\mu\nu}$ and $\kappa^2 = 8\pi G c^{-4} = 1/M^2_{pl}$. $M_{pl}$ is the (reduced) 
Planck mass $\sim 2 \times 10^{18}$ GeV. $S_m$ is the matter action that is 
independent of the connection but depends on the metric $g_{\mu\nu}$ and the 
matter field $\Phi$. Here, we will apply the Palatini formalism \cite{olmo, jyatsna}. 
As mentioned earlier, unlike the metric formalism, in the Palatini approach the 
torsion-free connection $\Gamma_{\mu\nu}^\alpha$ and the metric $g_{\mu\nu}$ 
are treated as the dynamical variables to be independently varied. Now, 
varying the action \eqref{eqn.1} with respect to the metric $g_{\mu\nu}$ we 
obtain the field equations as
\begin{equation}
f_\mathcal{R}(\mathcal{R})\, \mathcal{R}_{\mu\nu} - \frac{1}{2}f(\mathcal{R})\,g_{\mu\nu} = \kappa^2\, T_{\mu\nu},
\label{eqn.2}
\end{equation}
where $f_\mathcal{R}(\mathcal{R})$ is the derivative of $f(\mathcal{R})$ with respect to 
$\mathcal{R}$ and the energy-momentum tensor $T_{\mu\nu}$ is given by
\begin{equation}
T_{\mu\nu}\left[g^{\mu\nu}, \Phi\right] = \frac{-\,2}{\sqrt{-\,g}} \frac{\delta\left(\sqrt{-\,g}\,S_m\left[g^{\mu\nu}, \Phi\right]\right)}{\delta\, g_{\mu\nu}}.
\label{eqn.3}
\end{equation}
Again, the variation of the action \eqref{eqn.1} with respect to the 
connection $\Gamma_{\mu\nu}^\alpha$ gives,
\begin{equation}
\nabla_\alpha\! \left(\sqrt{-\,g}\, f_\mathcal{R}(\mathcal{R})\, g^{\mu\nu}\right) = 0.
\label{eqn.4}
\end{equation}
It should be noted that when $f(\mathcal{R}) = \mathcal{R}$, equation \eqref{eqn.2} gives the 
Einstein's field equations and equation \eqref{eqn.4} simply becomes the 
definition of the Levi-Civita connection in GR. This means that in the limit 
$f(\mathcal{R}) = \mathcal{R}$, the Palatini approach leads to GR as expected. Trace of equation 
\eqref{eqn.2} is given as
\begin{equation}
\mathcal{R}f_\mathcal{R}(\mathcal{R}) - 2f(\mathcal{R}) = \kappa^2\, T_\mu^\mu.
\label{eqn.5}
\end{equation}
This expression shows that $\mathcal{R} = g^{\mu\nu} \mathcal{R}_{\mu\nu}(\Gamma)$ can be 
algebraically solved in terms of the trace of energy-momentum tensor 
$T\equiv T_\mu^\mu = g^{\mu\nu} T_{\mu\nu}$ which leads to $\mathcal{R} = \mathcal{R}(T)$ and 
$f_\mathcal{R} = f_\mathcal{R}(T)$ as functions of matter but not of the torsion-free connection.

\subsection{Weyl transformation}
Named after Hermann Weyl, the Weyl transformation \cite{darabi, olmo} is a 
frame transformation from the Jordan frame to the Einstein frame. Usually 
this transformation is essential to get the minimally coupled scalar degree of 
freedom in the Einstein frame from the conventionally non-minimally coupled 
one in the Jordan frame in $f(\mathcal{R})$ gravity \cite{goswami}. Moreover, in the
Einstein frame the field equations in $f(\mathcal{R})$ gravity can be conveniently 
written in the form of GR.  In this transformation the Jordan frame metric 
$g_{\mu\nu}$ is related to the Einstein frame metric $\tilde g_{\mu\nu}$ as 
\begin{equation}
\tilde g_{\mu\nu} = \Omega\, g_{\mu\nu},
\label{eqn.6}
\end{equation}
where $\Omega$ is the conformal factor and in the present context it is simply 
given as $\Omega = f_\mathcal{R}(\mathcal{R})$. As a result, from equation \eqref{eqn.4} it can be 
noted that the connection for the torsion-free theory is the Levi-Civita 
connection for the conformally related metric $f_\mathcal{R}(\mathcal{R})\,g_{\mu\nu}$ 
\cite{olmo}. At this point it would be appropriate to recast equations
\eqref{eqn.2} for the torsion-free situation as 
\begin{equation}
\mathcal{R}_{\mu\nu} - \frac{1}{2}\mathcal{R}\,g_{\mu\nu} = \kappa^2\,\frac{T_{\mu\nu}}{f_\mathcal{R}(\mathcal{R})} - 
\frac{1}{2}\left(\mathcal{R} - \frac{f(\mathcal{R})}{f_\mathcal{R}(\mathcal{R})}\right)g_{\mu\nu}.
\label{eqn.7}
\end{equation}
However, in the case of torsion related environments where the connection is
asymmetric in nature, the above equation will take a different form containing
higher derivatives of $f(\mathcal{R})$, which can be noticed in the Ref.~\cite{olmo}. In
relation to this it should be mentioned that we consider here the torsion-free 
Palatini approach based on the fact that our considered model is the dark 
energy $f(\mathcal{R})$ gravity model satisfying the solar system tests and hence the 
model satisfies $f_\mathcal{R}(\mathcal{R}) - 1 \ll 1$ and also can be seen that 
$f_{\mathcal{RR}}(\mathcal{R}) \approx 0$ \cite{dhruba}. If we viewed the second term on the 
right hand side of equations \eqref{eqn.7} as the term for the effective 
cosmological constant $\Lambda_{e}$, i.e.
\begin{equation}
\Lambda_{e} = \frac{1}{2}\left(\mathcal{R} - \frac{f(\mathcal{R})}{f_\mathcal{R}(\mathcal{R})}\right)
\label{eqn.7a}
\end{equation}
in a spacetime having mass-energy source then the equations can be rewritten 
as the modified Einstein's field equations in $f(R)$ gravity in the form:
\begin{equation}
G_{\mu\nu} = \kappa^2\frac{T_{\mu\nu}}{f_\mathcal{R}(\mathcal{R})} - \Lambda_{e}\,g_{\mu\nu},
\label{eqn.8}
\end{equation}   
where $G_{\mu\nu}$ is the usual Einstein tensor.

It is clear that the Palatini field equations \eqref{eqn.8} are not in the 
complete form of Einstein's field equations in GR. Now through the 
conformal transformation \eqref{eqn.6} this can be achieved by transforming 
equations \eqref{eqn.8} from the Jordan frame to the Einstein frame as
\begin{equation}
\tilde G_{\mu\nu} = \kappa^2 \tilde T_{\mu\nu} - \tilde{\Lambda}_{e}\,\tilde g_{\mu\nu},
\label{eqn.9}
\end{equation}
where $\tilde G_{\mu\nu} = \mathcal{R}_{\mu\nu} - 1/2 \mathcal{R}\, \tilde g_{\mu\nu}$ is the 
Einstein tensor, $\tilde T_{\mu\nu} = T_{\mu\nu}/f_\mathcal{R}(\mathcal{R})$ is the 
energy-momentum tensor and 
\begin{equation}
\tilde{\Lambda}_e = \frac{1}{2}\left(\frac{\mathcal{R}}{f_\mathcal{R}(\mathcal{R})} - \frac{f(\mathcal{R})}{f_\mathcal{R}(\mathcal{R})^2}\right) 
\label{eqn.9a}
\end{equation}
is the effective cosmological constant in the Einstein frame. Thus the field
equations of $f(\mathcal{R})$ gravity in Palatini formalism in the Einstein frame can be
written exactly in the form of the corresponding equations in GR with the 
conformally transformed metric $\tilde{g}_{\mu\nu}$. Further, it is clear 
from equations \eqref{eqn.7a} and \eqref{eqn.9a} that the effective 
cosmological constant in the Einstein frame is different from that in the 
Jordan frame depending on the conformal factor $\Omega = f_\mathcal{R}(\mathcal{R})$.

\section{Rotational velocity of test particles around galaxies}
\label{sec.3}

In order to proceed towards the outcome of our work, we first need to consider 
a test particle (say, a star) in a galaxy moving in a stable circular 
orbit \cite{binney}. The centripetal acceleration $a$ of this test particle 
is related to its orbital velocity $v$ as
\begin{equation}
a = - \frac{v^2}{r}.
\label{eqn.10}
\end{equation}
Again, as the Einstein's equivalence principle can be justified for a theory 
of gravity that is conformally related to standard GR, hence the test particle 
in our case will satisfy the geodesic equation,
\begin{equation}
\frac{d^2 x^\mu}{ds^2} + \Gamma_{\rho\sigma}^\mu \frac{dx^\rho}{ds}\frac{dx^\sigma}{ds} = 0.
\label{eqn.11}
\end{equation}
To relate the orbital velocity of the test particle or star with its geodesic
motion we need to consider that although at the center of a galaxy the 
velocities of stars are very high, however it turns out that these velocities 
are very low in comparison to the speed of light. So the condition $v \ll c$ 
is always satisfied for the motion of stars in the galactic environment. Under 
this condition, using the coordinates $x^i = (ct, r, \theta, \phi)$ with 
$i = 0, 1, 2, 3$, it follows that
\begin{equation}
v^i = \left(\frac{dr}{dt},\; r\, \frac{d\theta}{dt},\; r \sin\theta\, \frac{d\phi}{dt} \right) \ll \frac{dx^0}{dt},
\label{eqn.12}
\end{equation}
where $x^0 = ct$.
Now, considering the above condition along with the weak field limit of 
equation \eqref{eqn.11} and a static spacetime ($\Gamma_{00}^0 = 0$), we get 
for the radial component,
\begin{equation}
\frac{d^2 r}{dt^2} = -\, c^2\, \Gamma_{00}^r.
\label{eqn.13}
\end{equation}
Thus from equations \eqref{eqn.10} and \eqref{eqn.13}, we obtain
\begin{equation}
v^2 (r) = r c^2\, \Gamma_{00}^r.
\label{eqn.14}
\end{equation}

It needs to be mentioned here that in the process of obtaining equation 
\eqref{eqn.4}, the idea of the invariance under the projective transformation
of the connection has been implemented \cite{sandberg, misner}. The projective 
transformation of the connection is defined as a class of connections 
related to each other as
\begin{equation}
\tilde \Gamma_{\mu\nu}^\lambda = \Gamma_{\mu\nu}^\lambda + \frac{2}{3}\delta_\nu^\lambda S_{\sigma\mu}^\sigma,
\label{eqn.15}
\end{equation}
where $S_{\sigma\mu}^\sigma \equiv A_\mu$ is a vector which generates torsion, 
and hence is usually set to zero from the beginning in the present study. 
Moreover, the different connections can define the same geodesics but are 
parametrized in different ways. The choice of parametrization is related to 
the choice of metric. In the case of Weyl transformation, the connection is a 
Levi-Civita connection of the conformal metric $\tilde g_{\mu\nu}$. Hence, we 
can write the connection in equation \eqref{eqn.14} as \cite{olmo}
\begin{equation}
\Gamma_{00}^r = \frac{1}{2}\,\tilde g^{r\sigma} \left(\tilde g_{\sigma 0,0} + \tilde g_{0\sigma, 0} - \tilde g_{00, \sigma}\right) = -\,\frac{1}{2}\,\tilde g^{rj}\partial_j \tilde g_{00}.
\label{eqn.16}
\end{equation}
The static spherically symmetric metric in the region exterior to the galactic 
baryonic mass distribution is given by the following line element:
\begin{equation}
ds^2 = \tilde g_{\mu\nu} dx^\mu dx^\nu = -\, e^{2\Phi} c^2 dt^2 + e^{2\Lambda} dr^2 + r^2 d\theta^2 + r^2 \sin^2\theta\, d\phi^2,
\label{eqn.17}
\end{equation}
where the metric coefficients $\Phi$ and $\Lambda$ are functions of the radial 
coordinate $r$ only. For this metric equation \eqref{eqn.16} takes the form:
\begin{equation}
\Gamma_{00}^r = e^{2\Phi - 2\Lambda}\, \Phi^\prime,
\label{eqn.18}
\end{equation}
where the prime denotes the derivative with respect $r$.

In terms of MTGs, the gravitational field equations can be generalized in the 
form \cite{mimoso, wojnar}:
\begin{equation}
\sigma(\varphi) (G_{\mu\nu} + H_{\mu\nu}) = \kappa^2 T_{\mu\nu},
\label{eqn.19}
\end{equation}
where $\sigma(\varphi)$ is a coupling factor to gravity and $\varphi$ 
generally represents either curvature invariants or other fields, such as 
scalar fields, which adds to the dynamics of the theory. The 
additional tensor $H_{\mu\nu}$ added to the Einstein tensor $G_{\mu\nu}$ 
represents the geometrical modifications which appear as a result of MTGs. In
this generalization GR can be recovered as a particular case of MTGs by 
considering $H_{\mu\nu} = 0$ and $\sigma(\varphi) = 1$. Keeping this 
representation in mind, we can rewrite equation \eqref{eqn.9} as 
\begin{equation}
f_\mathcal{R}(\mathcal{R}) \left(\tilde G_{\mu\nu} + \tilde{\Lambda}_{e}\,\tilde g_{\mu\nu}\right) = \kappa^2 T_{\mu\nu}
\label{eqn.20}
\end{equation}
It is seen that equation \eqref{eqn.20} is comparable to equation 
\eqref{eqn.19} when $H_{\mu\nu} = \tilde{\Lambda}_{e}\,\tilde g_{\mu\nu}$ and
$\sigma(\varphi) = f_\mathcal{R}(\mathcal{R})$. Moreover, in our study we consider the 
pressureless ($p = 0 $) dust model of the Universe and hence we take the trace 
of the energy-momentum tensor $T = -\, \rho$. Thus from the field equations 
\eqref{eqn.20} with the metric equation \eqref{eqn.17} one can rewrite the 
connection \eqref{eqn.18} as
\begin{equation}
\Gamma_{00}^r = \frac{e^{2(\Phi - \Lambda)}}{2\Omega} \left[\frac{\Omega(e^{2\Lambda} - 1)}{r} + 
\frac{\kappa^2 \rho}{\Omega}\, re^{2\Lambda} - e^{2\Lambda}\, r H_r^r\right].
\label{eqn.21}
\end{equation}
Now it is required to have the explicit form of the metric coefficients 
$e^{2\Phi}$ and $e^{2\Lambda}$ in terms of the galactic parameters through 
the $f(\mathcal{R})$ gravity model. Hence from the metric equation 
\eqref{eqn.17}, following the thorough calculations \cite{wojnar}, we are 
able to obtain the metric coefficient $e^{2\Lambda}$ in the familiar form as 
given by
\begin{equation}
e^{2\Lambda} = \left(1 - \frac{2GM'(r)}{c^2 r}\right)^{-1}\!\!\!\!,
\label{eqn.22}
\end{equation}
where the modified mass distribution is defined as
\begin{equation}
M^\prime(r) = \frac{M(r)}{\Omega} = \frac{1}{\Omega} \int_0^r \left(\frac{\kappa^2 \rho}{2 G c^2} \frac{r^2}{f_\mathcal{R}(r)} - \Omega \frac{c^2 r^2 H_{tt}(r)}{2G e^{2\Phi}}\right)dr.
\label{eqn.23}
\end{equation}
Whereas the exact form of the metric coefficient $e^{2\Phi}$ is very complex. 
Hence, we employ another way of solving it as follows. It is reasonable to 
assume that the metric coefficients $e^{2\Phi}$ and $e^{2\Lambda}$ take the 
Schwarzschild form at large distances compared to the core radius of a galaxy. 
This form of the coefficient $e^{2\Lambda}$ is already seen in equation 
\eqref{eqn.22}. Hence, from the Ref.~\cite{Weinberg} we can 
write the metric $\tilde{g}_{\mu\nu}$ for the weak field approximation as
\begin{equation}
\tilde g_{\mu\nu} = \Omega\, (\eta_{\mu\nu} + g_{\mu\nu}),
\label{eqn.24}
\end{equation}
where $\eta_{\mu\nu}$ is the metric of the Minkowski spacetime and the first 
order term coming from the Newtonian limit is given as
\begin{equation}
g_{00} \approx -\, \frac{2\phi}{c^2} = \frac{2GM'(r)}{c^2 r}.
\label{eqn.25}
\end{equation}
Thus from equation \eqref{eqn.24}, we can write,
\begin{equation}
e^{2\Phi} \approx \Omega\, (1 - g_{00}) = 1 - \frac{2GM(r)}{\Omega c^2 r}.
\label{eqn.26}
\end{equation}

As we intend to study the behaviour of rotation curves of test particles in a 
galaxy moving in a stable circular orbit, we require to find an exact form of 
the connection \eqref{eqn.21} for the recently introduced model
\cite{dhruba} of $f(\mathcal{R})$ gravity.

\section{Rotational velocity in the new $f(\mathcal{R})$ gravity model}
\label{sec.4}

In the new $f(\mathcal{R})$ gravity model \cite{dhruba, Dhruba},
\begin{equation}
f(\mathcal{R}) = \mathcal{R} - \frac{\alpha}{\pi} \mathcal{R}_c \cot^{-1} \left(\frac{\mathcal{R}_c^2}{\mathcal{R}^2}\right) - 
\beta \mathcal{R}_c \left[1 - \exp\left(- \frac{\mathcal{R}}{\mathcal{R}_c}\right)\right],
\label{eqn.27}
\end{equation}
where $\alpha$ and $\beta$ are two dimensionless constants. $\mathcal{R}_c$ is 
the characteristic curvature constant with dimension similar to curvature 
scalar $\mathcal{R}$, and one may expect $\mathcal{R}_c \backsim \Lambda$. For 
this model, the conformal factor is obtained as
\begin{equation}
\Omega = f_\mathcal{R}(\mathcal{R}) = 1 - \frac{2\,\alpha \mathcal{R}_c^3}{\pi \mathcal{R}^3 \left(\frac{\mathcal{R}_c^4}{\mathcal{R}^4} + 1\right)} - \beta \exp \left(- \frac{\mathcal{R}}{\mathcal{R}_c}\right).
\label{eqn.28}
\end{equation}
Using this conformal factor, and equations \eqref{eqn.22} and \eqref{eqn.26}  
we are able to derive the exact form of the connection \eqref{eqn.21} for the 
aforementioned modified gravity model as
\begin{equation}
\Gamma_{00}^r = \frac{GM}{\Omega c^2 r^2} \Bigg[1 - \frac{2GM}{\Omega c^2 r} + r^2\left(\frac{\Omega c^2 r}{2GM} - 1\right)\Bigg\{\!\!\left(1 - \frac{1}{2\Omega}\right)\left(\frac{2f(\mathcal{R}) - \mathcal{R}f_\mathcal{R}(\mathcal{R})}{\Omega^2}\right) + \frac{f(\mathcal{R})}{2\Omega^3}\Bigg\}\Bigg].
\label{eqn.29}
\end{equation}
Finally, we arrive at the expression of rotational velocity for the 
new model \eqref{eqn.27} from equation \eqref{eqn.14} as
\begin{equation}
v^2(r) = \frac{GM}{f_\mathcal{R}(\mathcal{R}) r} \Bigg[1 - \frac{2GM}{f_\mathcal{R}(\mathcal{R}) c^2 r} + r^2\left(\frac{f_\mathcal{R}(\mathcal{R}) c^2 r}{2GM} - 1\right)\Bigg\{\!\!\left(1 - \frac{1}{2f_\mathcal{R}(\mathcal{R})}\right)\left(\frac{2f(\mathcal{R}) - \mathcal{R}f_\mathcal{R}(\mathcal{R})}{f_\mathcal{R}(\mathcal{R})^2}\right) + \frac{f(\mathcal{R})}{2f_\mathcal{R}(\mathcal{R})^3}\Bigg\}\Bigg].
\label{eqn.30}
\end{equation}
Here we should mention that because of the complicated form of the modified 
mass distribution equation \eqref{eqn.23}, in our work we assume a simple mass 
distribution equation within a galaxy in the form \cite{sporea}:
\begin{equation}
M(r) = M_0 \left(\sqrt{\frac{R_0}{r_c}} \frac{r}{r + r_c}\right)^{3\gamma}\!\!\!,
\label{eqn.31}
\end{equation}
where $M_0$ (total mass of the galaxy) and $r_c$ (core radius) are the 
parameters to be predicted by fitting the calculated rotational velocities 
with observed data. $R_0$ is the scale length of the galaxy. And, the values 
of the parameter $\gamma$ \cite{hybrid_metric_palatini} are $\gamma = 1$ for 
high surface brightness (HSB) galaxies and $\gamma = 2$ for low surface 
brightness (LSB) and dwarf galaxies.

\begin{table}[h!]
\begin{center}
\caption{Relevant galaxy properties for a set of nine HSB galaxies and the 
best fit values for the parameters $M_0$ and $r_c$ of these galaxies in the 
new $f(\mathcal{R})$ gravity model \eqref{eqn.27}.}
\begin{tabular}{ c c c c c c c c c }
\hline \hline 
Galaxy & Distance & $L_B$ & $R_0 $ & $M_{HI}$ & $M_0$ & $r_c$ & $\chi_{red}^2$ & $(M/L)_\ast$ \\ \vspace{0.5mm}
  & $D$ (Mpc) & $(10^{10}L_\odot)$ & (kpc) & $(10^{10} M_\odot)$ & $(10^{10}M_\odot)$ & (kpc) &  & $({M_\odot}/{L_\odot})$ \\ 
   \hline 
 NGC 3877 & 15.5 & 1.948 & 2.4 & 0.11 & 89.878 & 5.53 & 0.82 & 46.06\\
 NGC 3893 & 18.1 & 2.928 & 2.4 & 0.59 & 32.387 & 3.36 & 0.47 & 10.79\\
 NGC 3949 & 18.4 & 2.327 & 1.7 & 0.35 & 27.868 & 2.97 & 0.09 & 11.77\\
 NGC 3953 & 18.7 & 4.236 & 3.9 & 0.31 & 67.509 & 5.36 & 0.08 & 15.84\\
 NGC 3972 & 18.6 & 0.978 & 2.0 & 0.13 & 97.868 & 6.20 & 0.27 & 99.89\\
 NGC 4183 & 16.7 & 1.042 & 2.9 & 0.30 & 28.603 & 5.54 & 0.27 & 27.07\\
 NGC 4217 & 19.6 & 3.031 & 3.1 & 0.30 & 50.422 & 4.77 & 0.85 & 16.50\\
 NGC 4389 & 15.5 & 0.610 & 1.2 & 0.04 & 1499.78 & 11.89 & 0.09 & 2458.57\\
 NGC 6946 & 6.9 & 3.732 & 2.9 & 0.57 & 24.578 & 3.61 & 15.13 & 6.38\\
 \hline \hline   
\end{tabular}
\label{Table:1}
\end{center}
\end{table}

\section{Astrophysical tests of the new ${f(\mathcal{R})}$ gravity model in galactic levels}
\label{sec.5}

In this section we check the validity of the new $f(\mathcal{R})$ gravity model \eqref{eqn.27} 
by comparing its theoretical predictions with the observational data on the galactic 
rotation curves. Using equation \eqref{eqn.30}, we try to obtain the flat 
rotation curves for a few samples of HSB, LSB and dwarf galaxies in the 
subsections that follow. We also investigate the rotation curve of an 
ultra diffuse galaxy (UDG) AGC 242019. For this analysis, we have used a well constrained 
set of model parameters viz. $\frac{\mathcal{R}}{\mathcal{R}_c} = 1.5$, $\alpha = 0.005$ and 
$\beta = 0.044$ \cite{dhruba}.

\begin{figure}[!h]
\includegraphics[scale = 0.28]{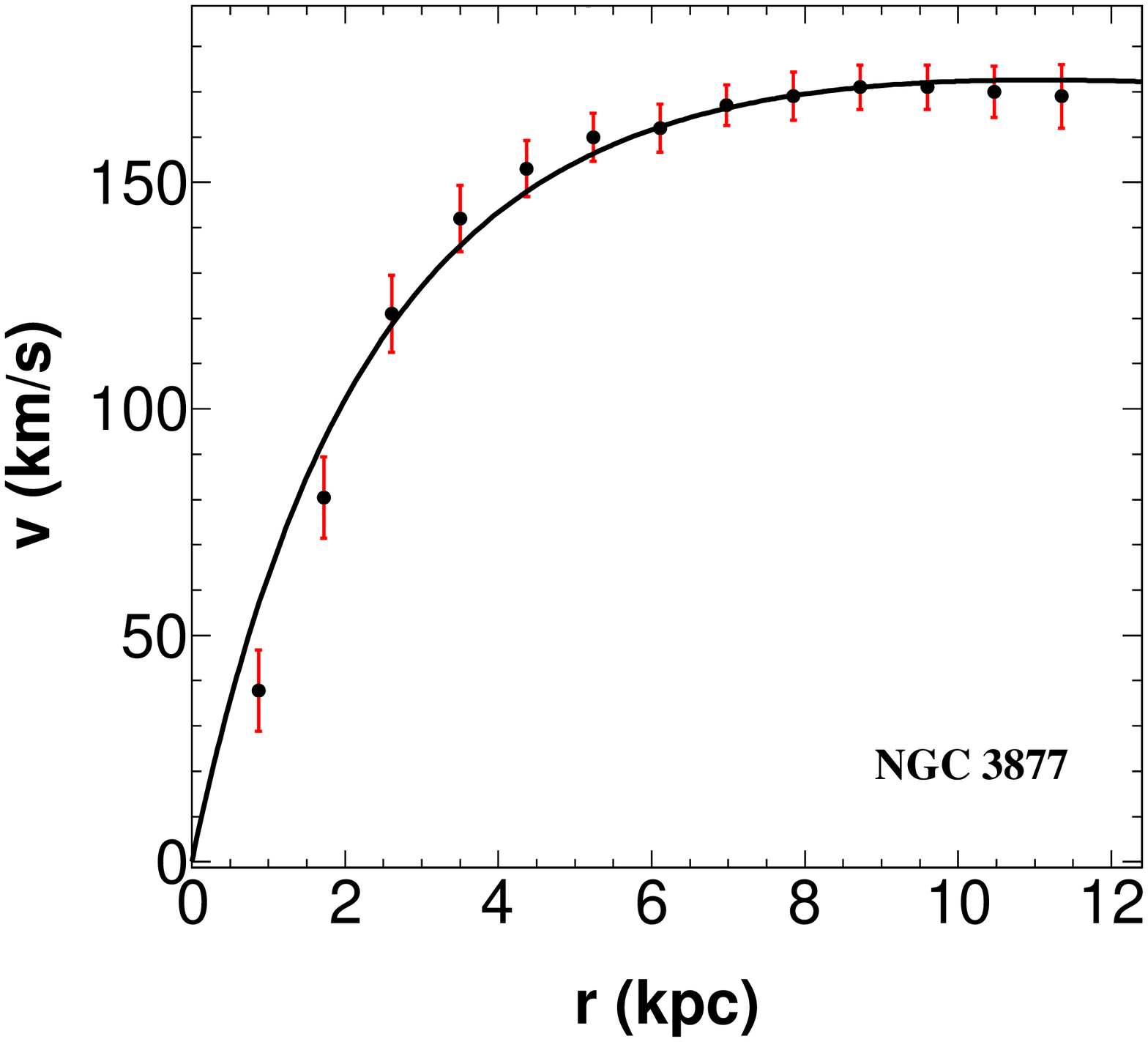} \hspace{0.2cm}
\includegraphics[scale = 0.28]{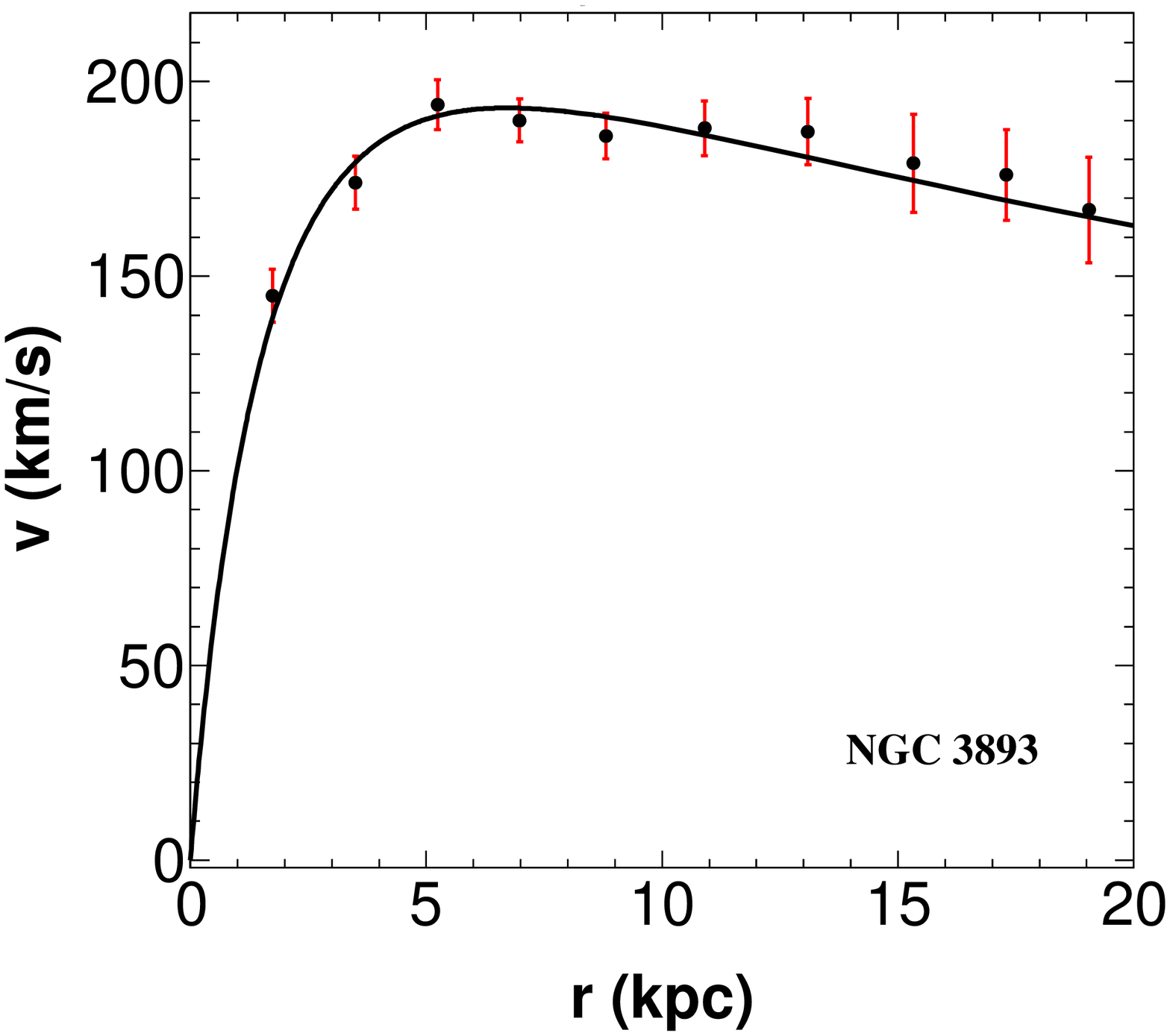} \hspace{0.2cm}
\includegraphics[scale = 0.28]{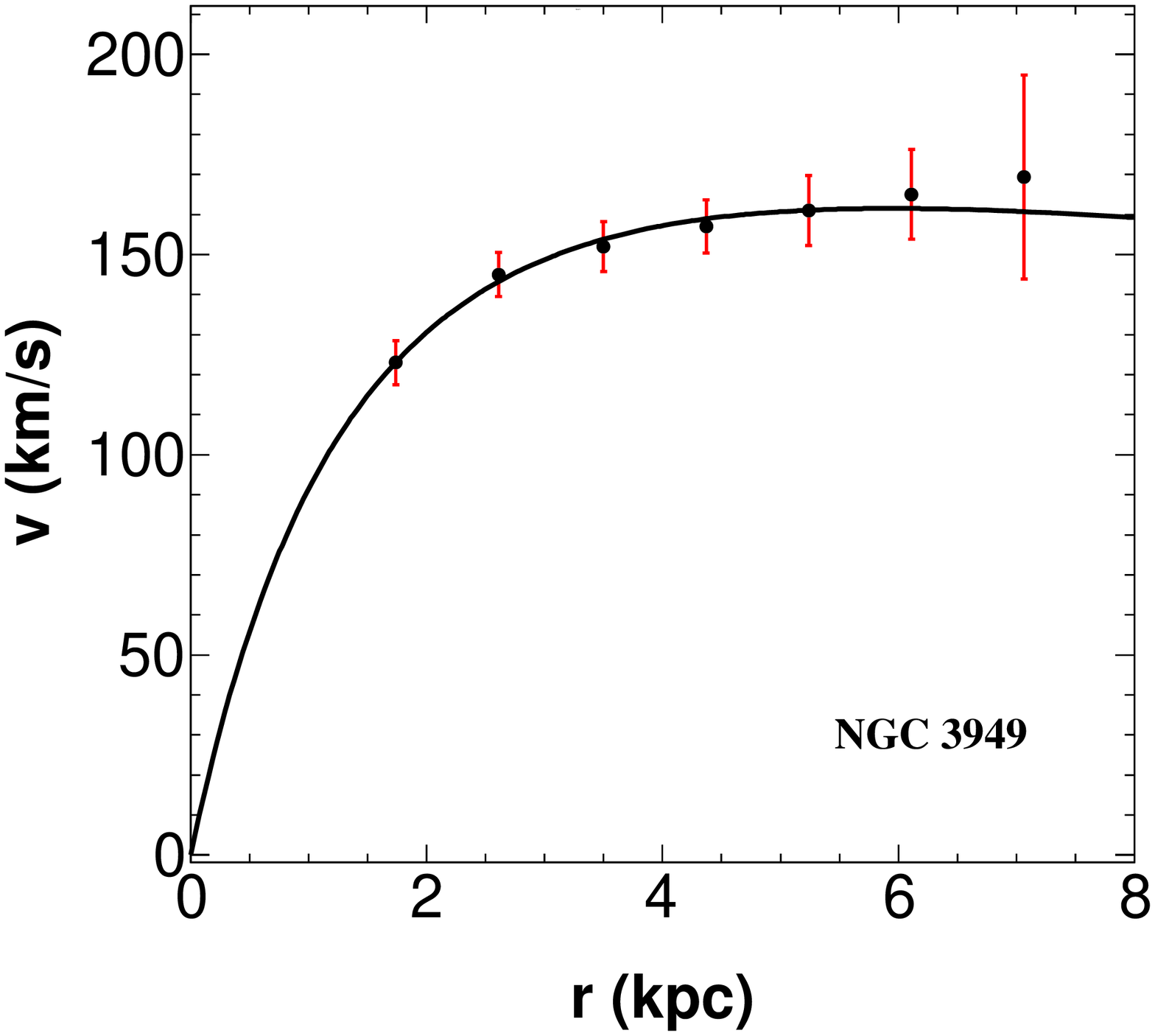} \hspace{0.2cm}
\includegraphics[scale = 0.28]{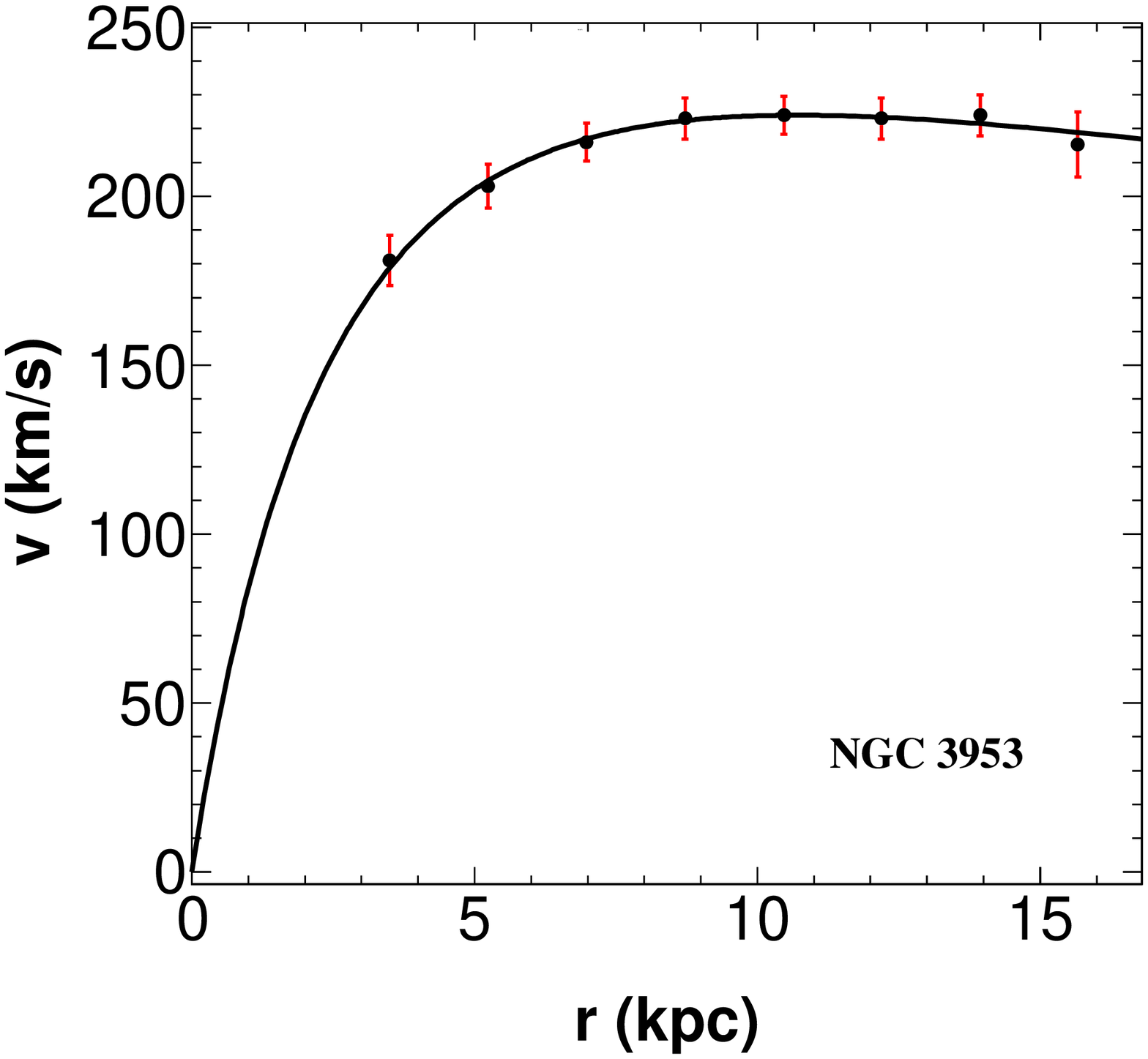} \hspace{0.2cm}
\includegraphics[scale = 0.28]{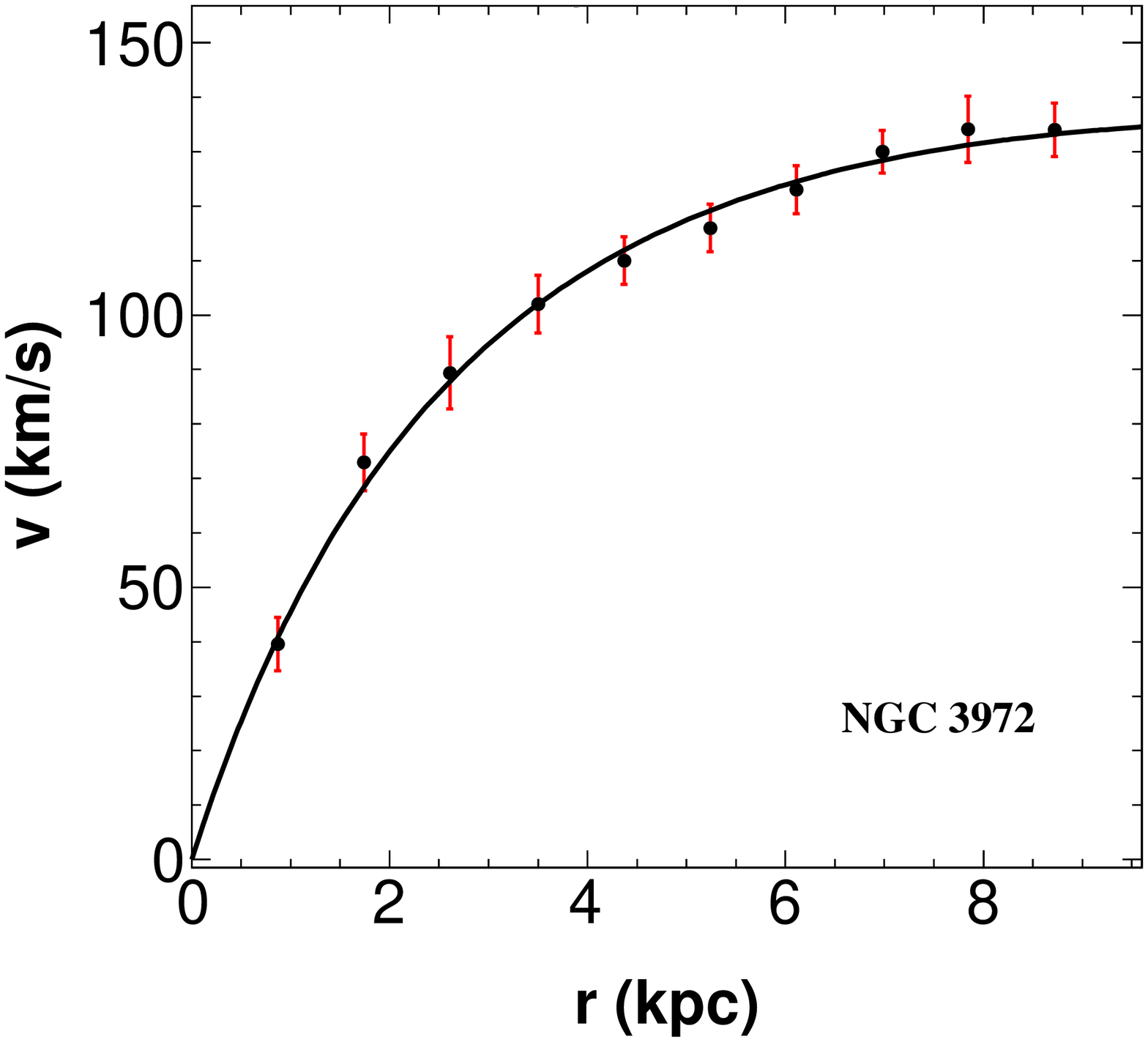} \hspace{0.2cm}
\includegraphics[scale = 0.28]{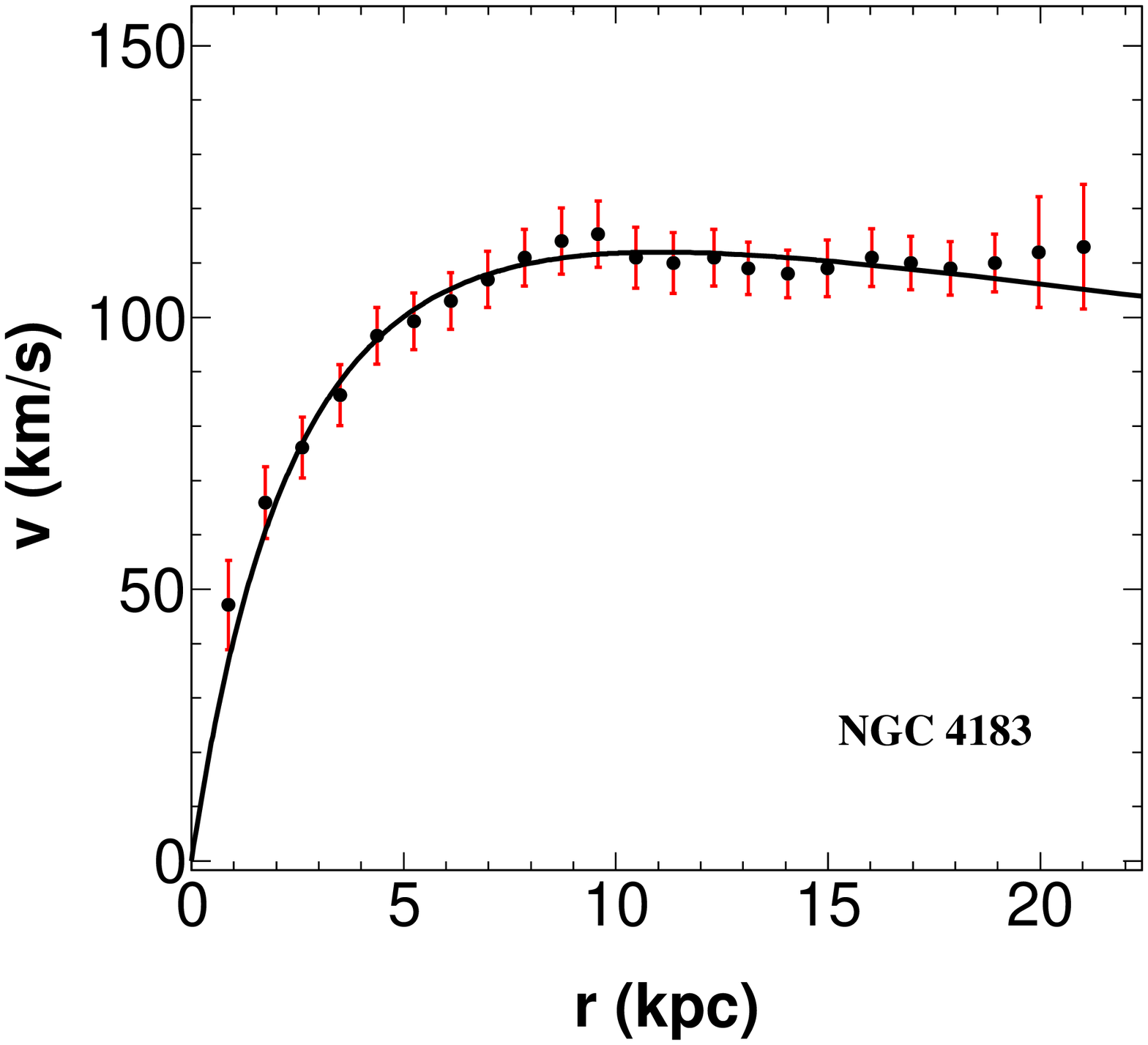} \hspace{0.2cm}
\includegraphics[scale = 0.28]{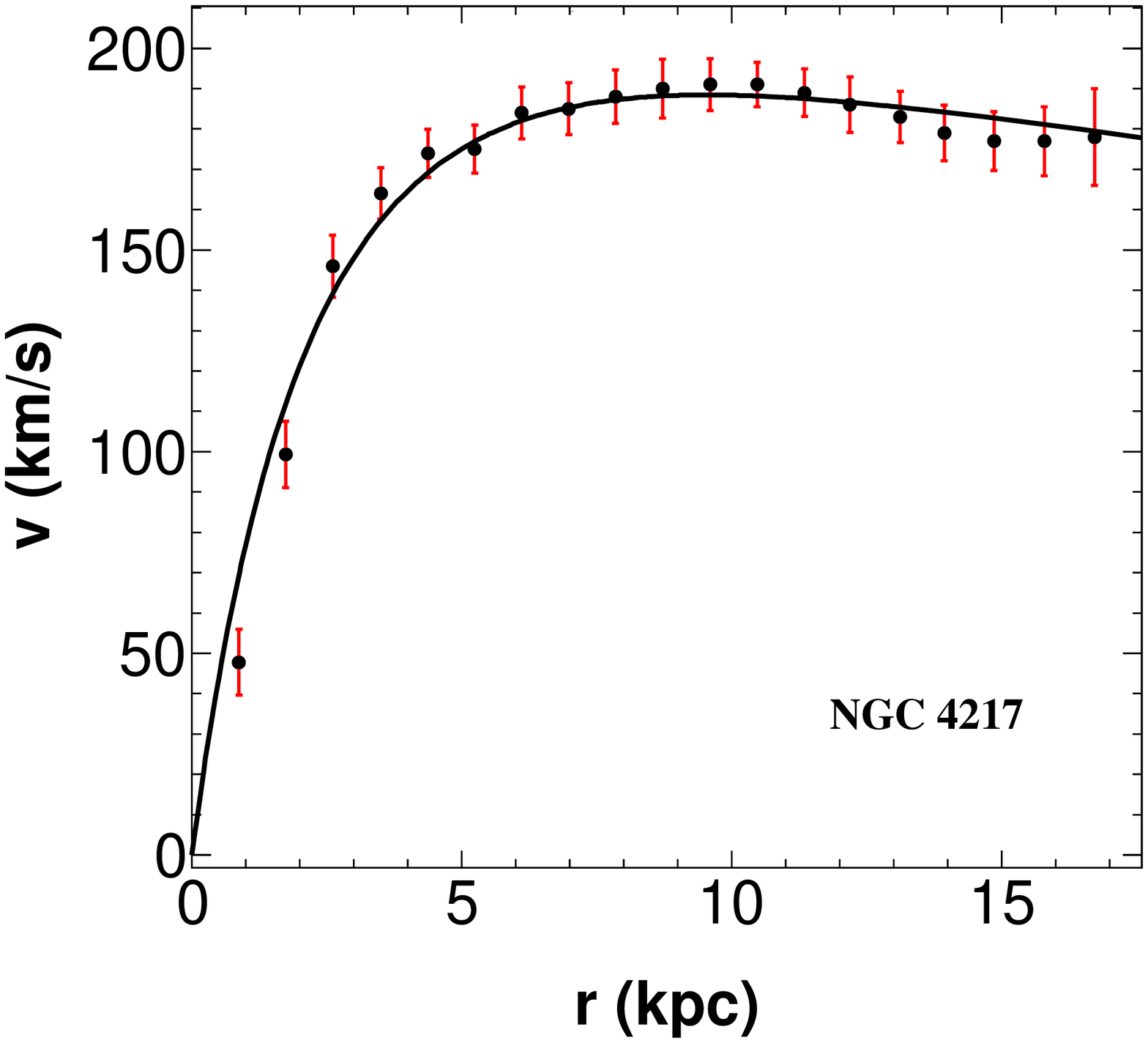} \hspace{0.2cm}
\includegraphics[scale = 0.28]{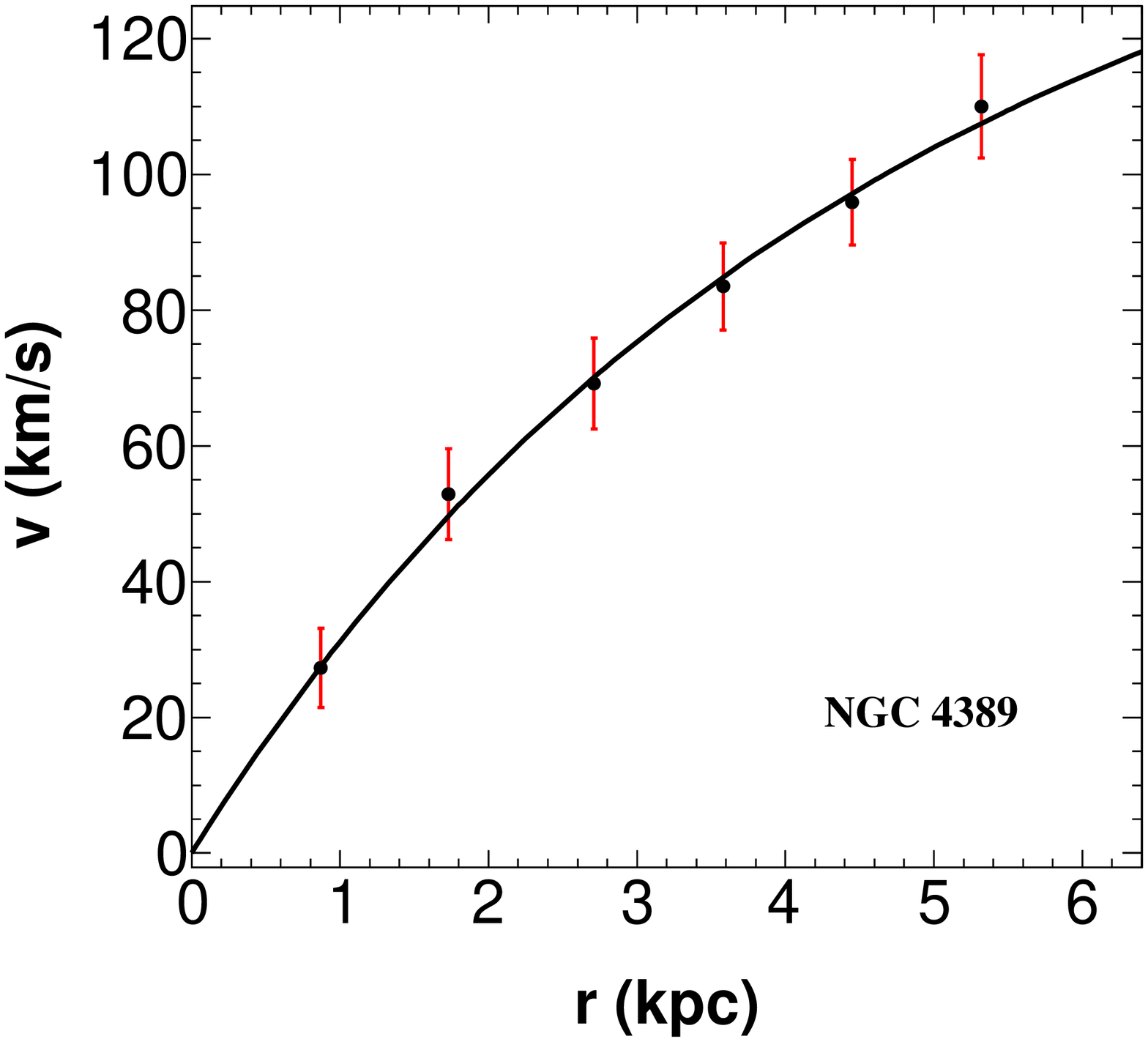} \hspace{0.2cm}
\includegraphics[scale = 0.28]{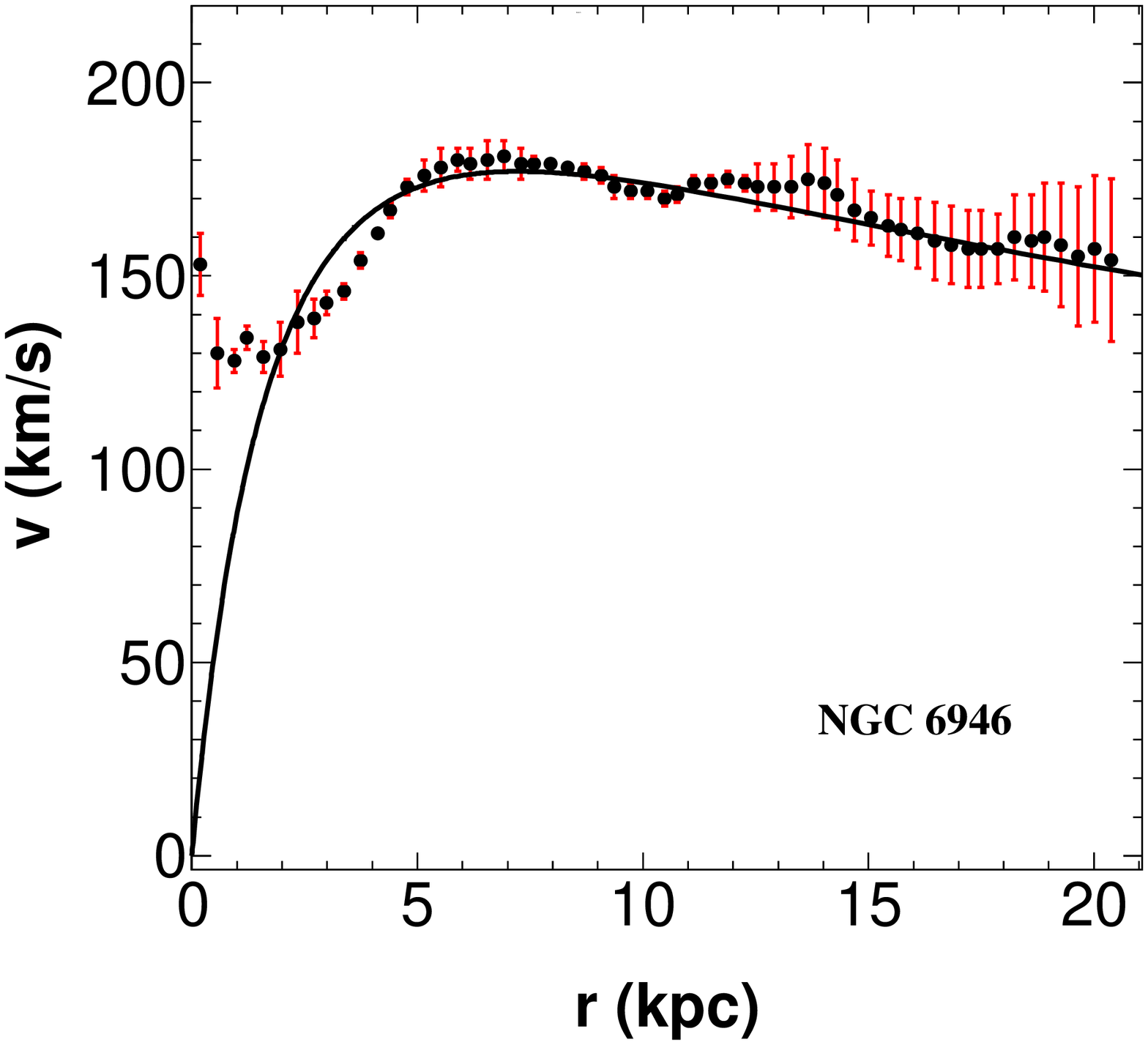} \hspace{0.2cm}
\caption{Fitting of equation \eqref{eqn.30} for the new $f(\mathcal{R})$ gravity model 
\eqref{eqn.27} to the rotational velocities (in km/s) of a set of nine HSB galaxies extracted from 
Ref.~\cite{mannheim} with their errors plotted as a function of radial 
distance (in kpc).}
\label{fig.1}
\end{figure}

\subsection{Analysis of HSB galaxy sample}

We fit the predictions of the new $f(\mathcal{R})$ gravity model \eqref{eqn.27} 
as obtained from equation \eqref{eqn.30} to the data of a sample of nine HSB galaxies extracted from 
Ref.~\cite{mannheim}. In these fittings and subsequent ones we use the $\chi^2$ 
minimization technique. Table \ref{Table:1} shows the related data for the 
nine HSB galaxies along with the respective best fit values of the parameters 
$M_0$ and $r_c$ and the reduced chi-squared ($\chi_{red}^2$) values. The 
results are depicted in Fig.\ \ref{fig.1}. 

In Table \ref{Table:1}, the data of HI gas mass $M_{HI}$ (which extends well 
beyond the optical disk distributions) in units of $10^{10} M_\odot$, the 
disk scale length $R_0$ measured in units of kpc, the adopted distance $D$ in 
Mpc and the B-band luminosity in units of $10^{10}L_\odot$ are extracted from 
Ref~\cite{mannheim}. The total gas mass is given by $\frac{4}{3} M_{HI}$ 
\cite{capistrano, sporea, mannheim, karukes} and the stellar mass is obtained 
by subtracting the gas mass from the 
predicted total mass of the galaxy. We have also listed the stellar 
mass-to-light ratios $(M/L)_\ast$. It is seen that in almost all the cases, 
the mass-to-light ratios inferred from the fit are reasonably close to the 
expected mass-to-light ratios. One should note that the expected mass-to-light 
ratios of galaxies lie within the range of $0.2\, M_\odot/L_\odot < M/L <
10\, M_\odot/L_\odot$ \cite{sanders}.  Also, Fig.\ \ref{fig.1} depicts that 
almost all of the selected galaxies are well-fitted by the recently proposed 
$f(\mathcal{R})$ model \eqref{eqn.27}, except the galaxy NGC $6946$ for which 
the $\chi_{red}^2$ value is very large, around $15.13$. However, as one can see 
from Fig.\ \ref{fig.1}, the shape of its rotation curve was not affected. 

\begin{table}[h!]
\begin{center}
\caption{Relevant galaxy properties for a set of 21 LSB galaxies and the best 
fit values for the parameters $M_0$ and $r_c$ of these galaxies in the
new $f(\mathcal{R})$ gravity model \eqref{eqn.27}.}
\begin{tabular}{ c c c c c c c c c }
\hline \hline 
Galaxy & Distance & $L_B$ & $R_0 $ & $M_{HI}$ & $M_0$ & $r_c$ & $\chi_{red}^2$ & $(M/L)_\ast$ \\ \vspace{0.5mm}
  & $D$ (Mpc) & $(10^{10}L_\odot)$ & (kpc) & $(10^{10} M_\odot)$ & $(10^{10}M_\odot)$ & (kpc) &  & $({M_\odot}/{L_\odot})$ \\ 
   \hline 
 DDO 0064 & 6.8 & 0.015 & 1.3 & 0.02 & 0.209 & 0.84 & 0.63 & 12.15\\
 F563-1 & 46.8 & 0.140 & 2.9 & 0.29 & 3.365 & 2.09 & 0.53 & 21.27\\
 F563-V2 & 57.8 & 0.266 & 2.0 & 0.20 & 1.917 & 1.34 & 0.12 & 6.20\\
 F568-3 & 80.0 & 0.351 & 4.2 & 0.30 & 2.100 & 2.66 & 0.38 & 4.84\\
 F583-1 & 32.4 & 0.064 & 1.6 & 0.18 & 14.398 & 2.24 & 0.34 & 221.21\\
 NGC 0055 & 1.9 & 0.588 & 1.9 & 0.13 & 10.976 & 2.35 & 1.22 & 16.01\\
 NGC 0300 & 2.0 & 0.271 & 2.1 & 0.08 & 2.394 & 1.65 & 2.04 & 8.44\\
 NGC 2976 & 3.6 & 0.201 & 1.2 & 0.01 & 0.945 & 0.84 & 1.42 & 4.63\\
 NGC 3109 & 1.5 & 0.064 & 1.3 & 0.06 & 3.410 & 1.52 & 2.78 & 52.03\\
 NGC 3917 & 16.9 & 1.334 & 2.8 & 0.17 & 7.774 & 2.26 & 0.42 & 5.65\\
 UGC 1281 & 5.1 & 0.017 & 1.6 & 0.03 & 0.950 & 1.34 & 0.13 & 53.54\\
 UGC 5005 & 51.4 & 0.200 & 4.6 & 0.28 & 13.277 & 4.43 & 0.34 & 64.52\\
 UGC 5750 & 56.1 & 0.472 & 3.3 & 0.10 & 9.703 & 3.56 & 0.07 & 20.27\\
 UGC 5999 & 44.9 & 0.170 & 4.4 & 0.18 & 6.468 & 3.57 & 0.09 & 36.64\\
 UGC 6399 & 18.7 & 0.291 & 2.4 & 0.07 & 1.269 & 1.62 & 0.21 & 4.04\\
 UGC 6667 & 19.8 & 0.422 & 3.1 & 0.10 & 0.548 & 1.60 & 0.65 & 0.98\\
 UGC 6818 & 21.7 & 0.352 & 2.1 & 0.16 & 8.677 & 2.46 & 1.69 & 24.04\\
 UGC 6923 & 18.0 & 0.297 & 1.5 & 0.08 & 1.005 & 1.11 & 0.65 & 3.02\\
 UGC 6983 & 20.2 & 0.577 & 2.9 & 0.37 & 1.287 & 1.67 & 1.11 & 1.37\\
 UGC 7089 & 13.9 & 0.352 & 2.3 & 0.07 & 2.013 & 1.92 & 0.85 & 5.45\\
 UGC 11557 & 23.7 & 1.806 & 3.0 & 0.25 & 2.043 & 2.18 & 0.18 & 0.95\\
 \hline \hline   
\end{tabular}
\label{Table:2}
\end{center}
\end{table}

\begin{figure}[!h]
\includegraphics[scale = 0.28]{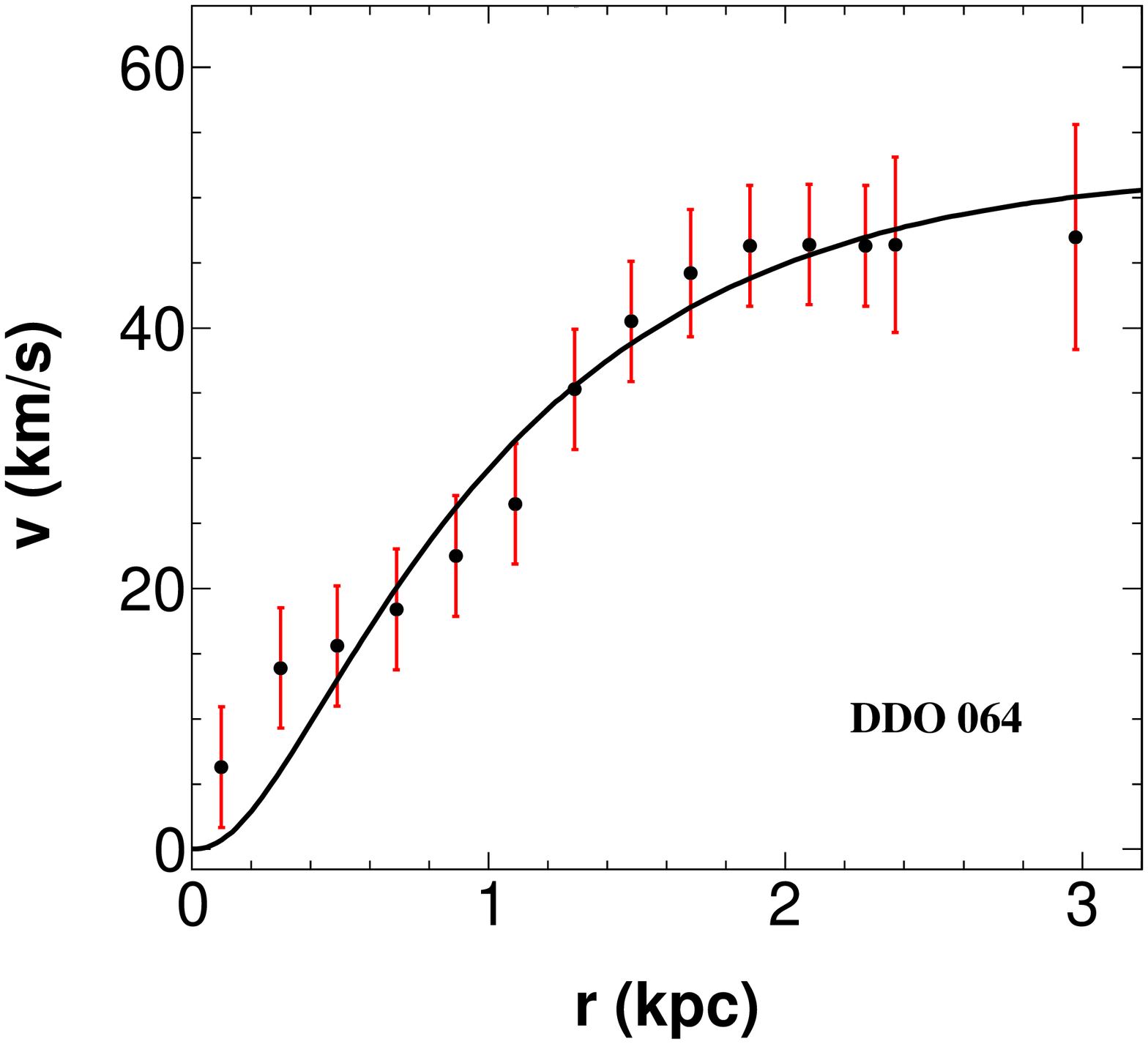} \hspace{0.2cm}
\includegraphics[scale = 0.28]{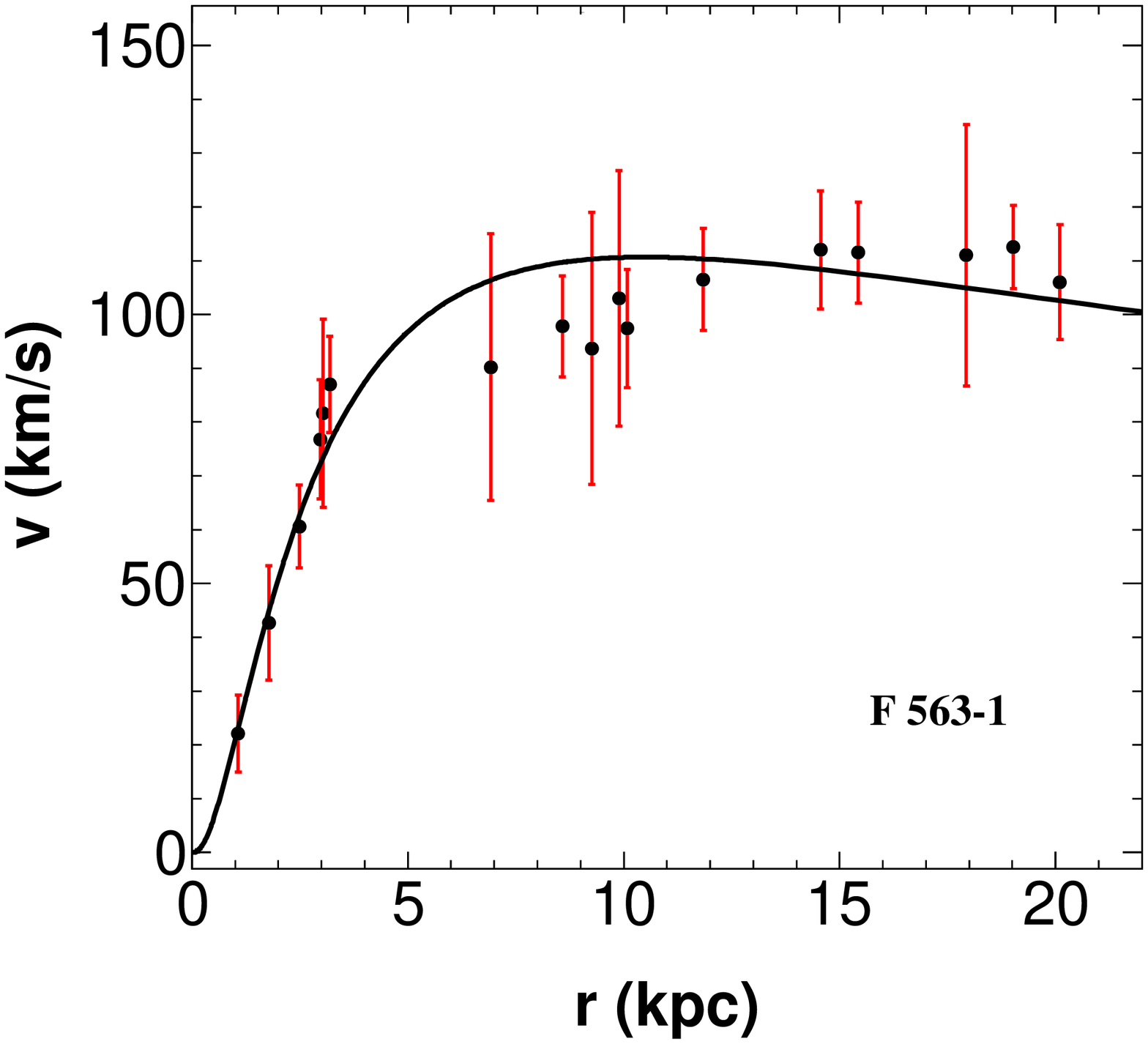} \hspace{0.2cm}
\includegraphics[scale = 0.28]{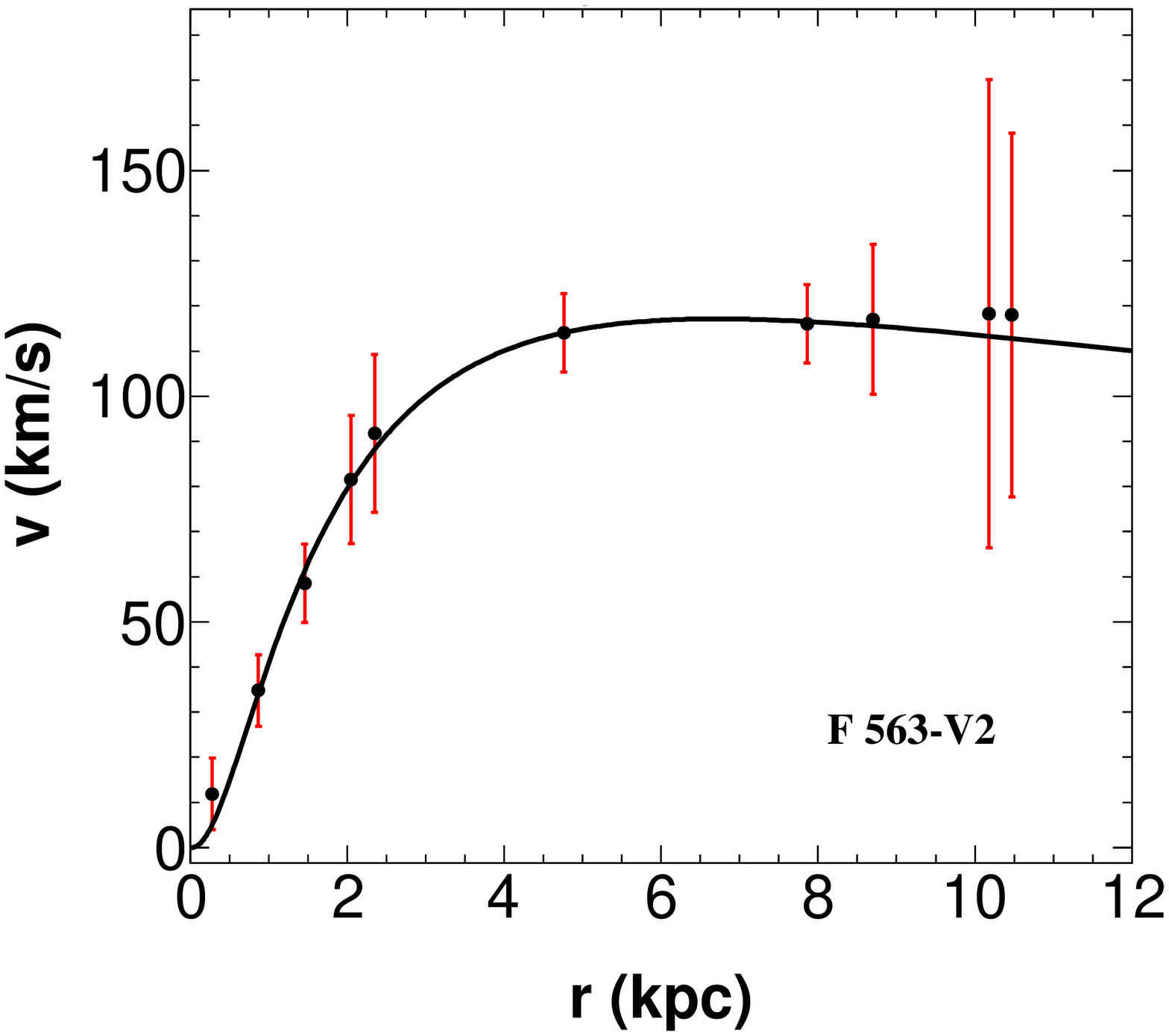} \hspace{0.2cm}
\includegraphics[scale = 0.28]{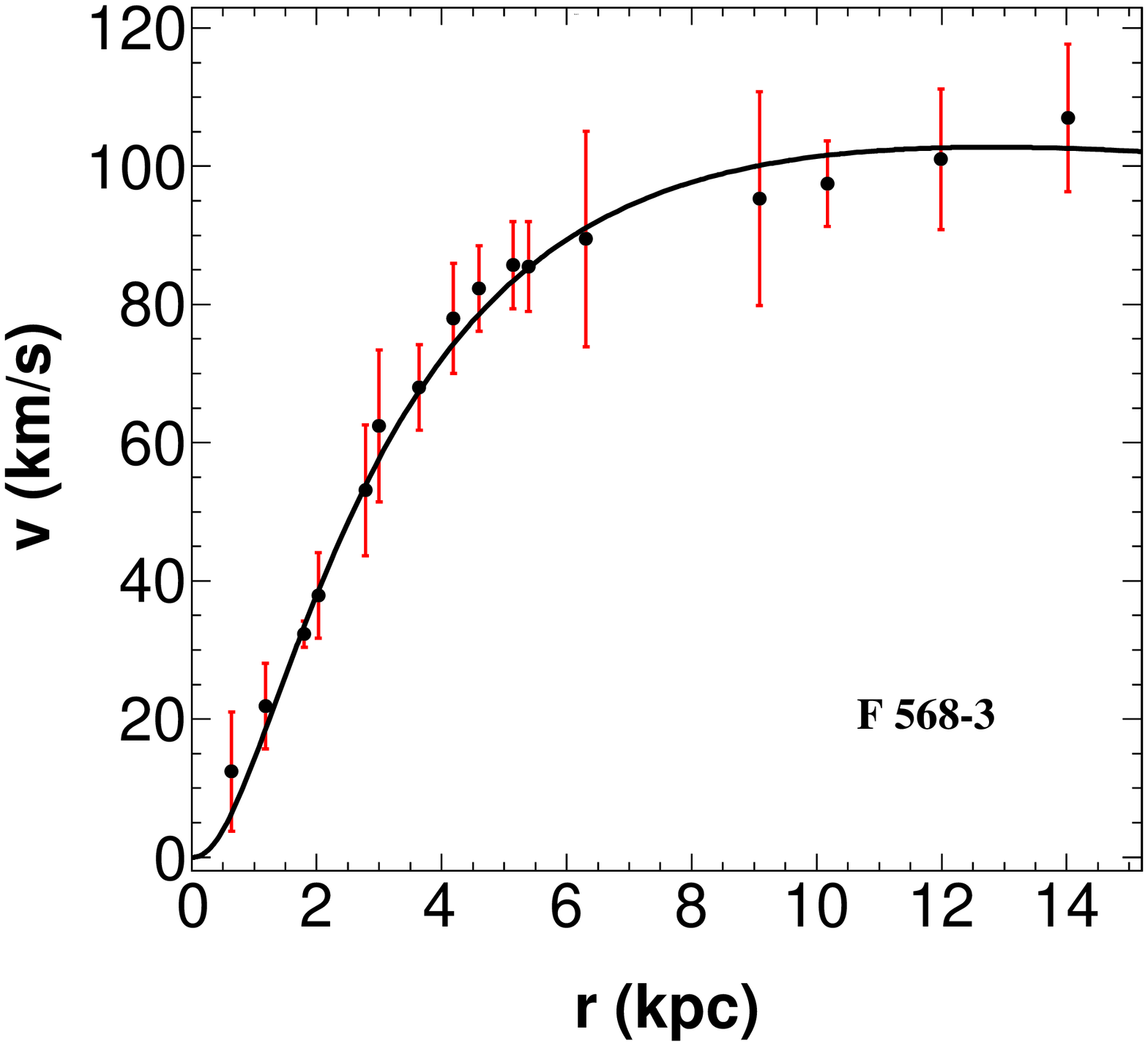} \hspace{0.2cm}
\includegraphics[scale = 0.28]{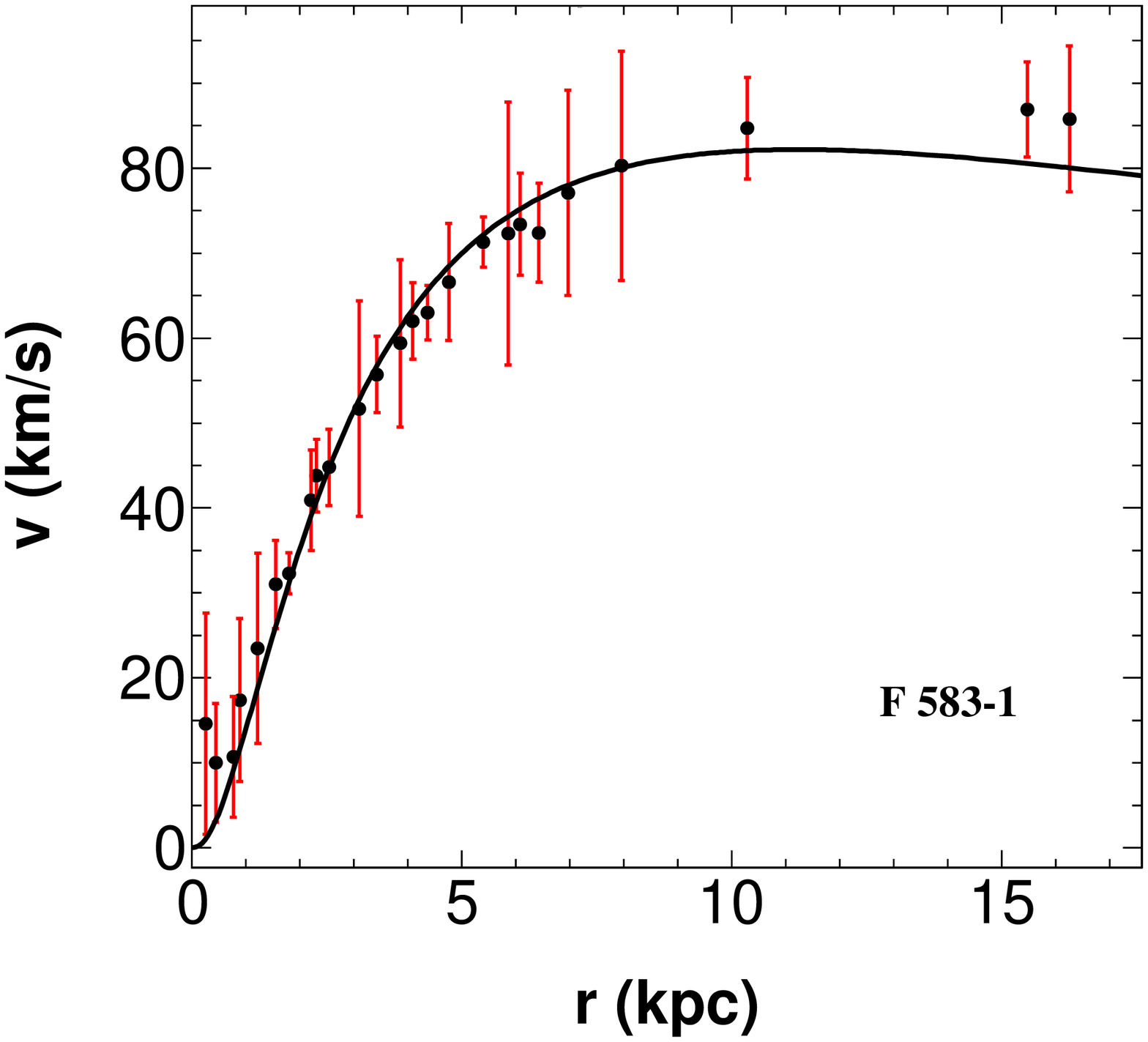} \hspace{0.2cm}
\includegraphics[scale = 0.28]{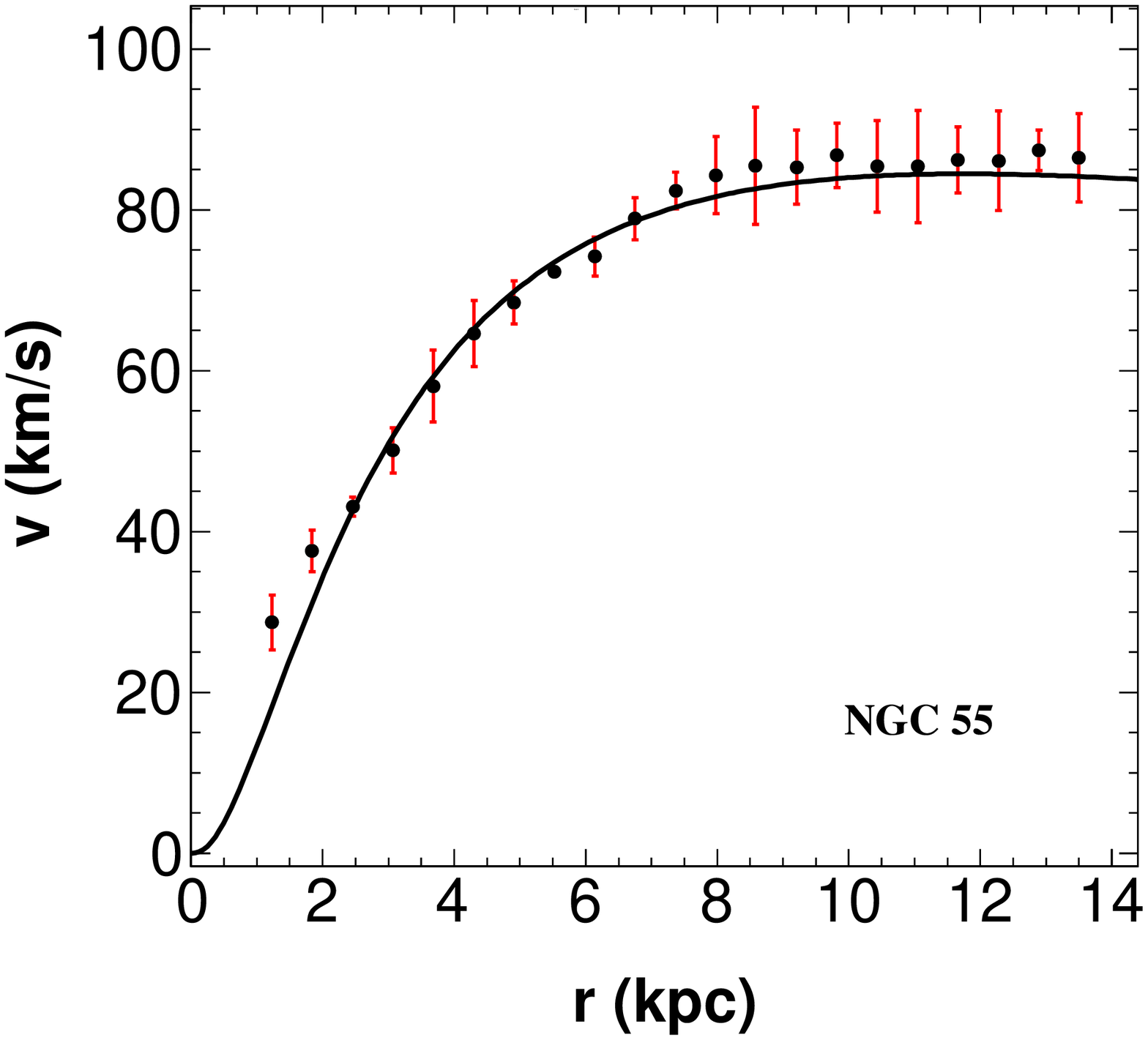} \hspace{0.2cm}
\includegraphics[scale = 0.28]{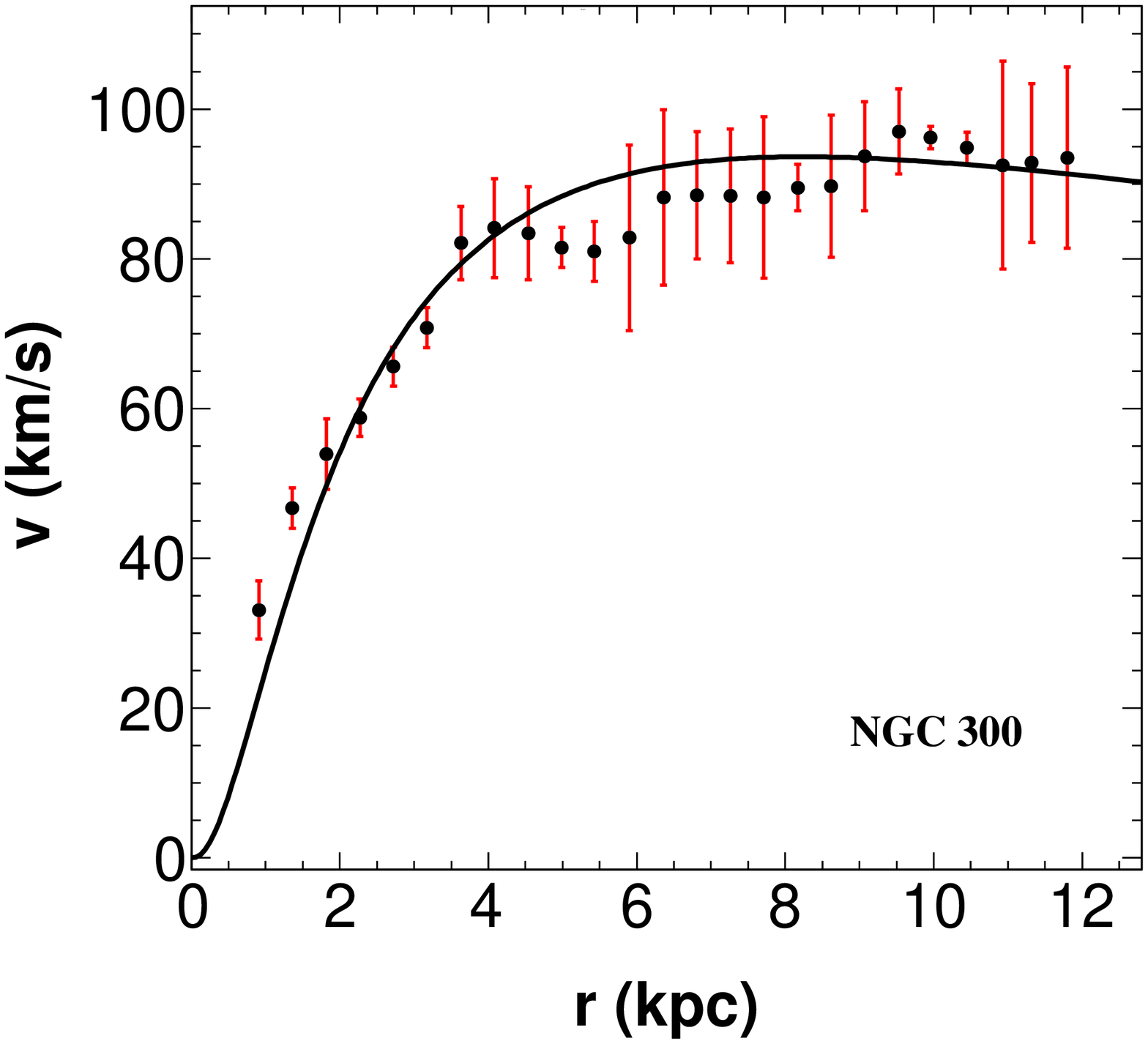} \hspace{0.2cm}
\includegraphics[scale = 0.28]{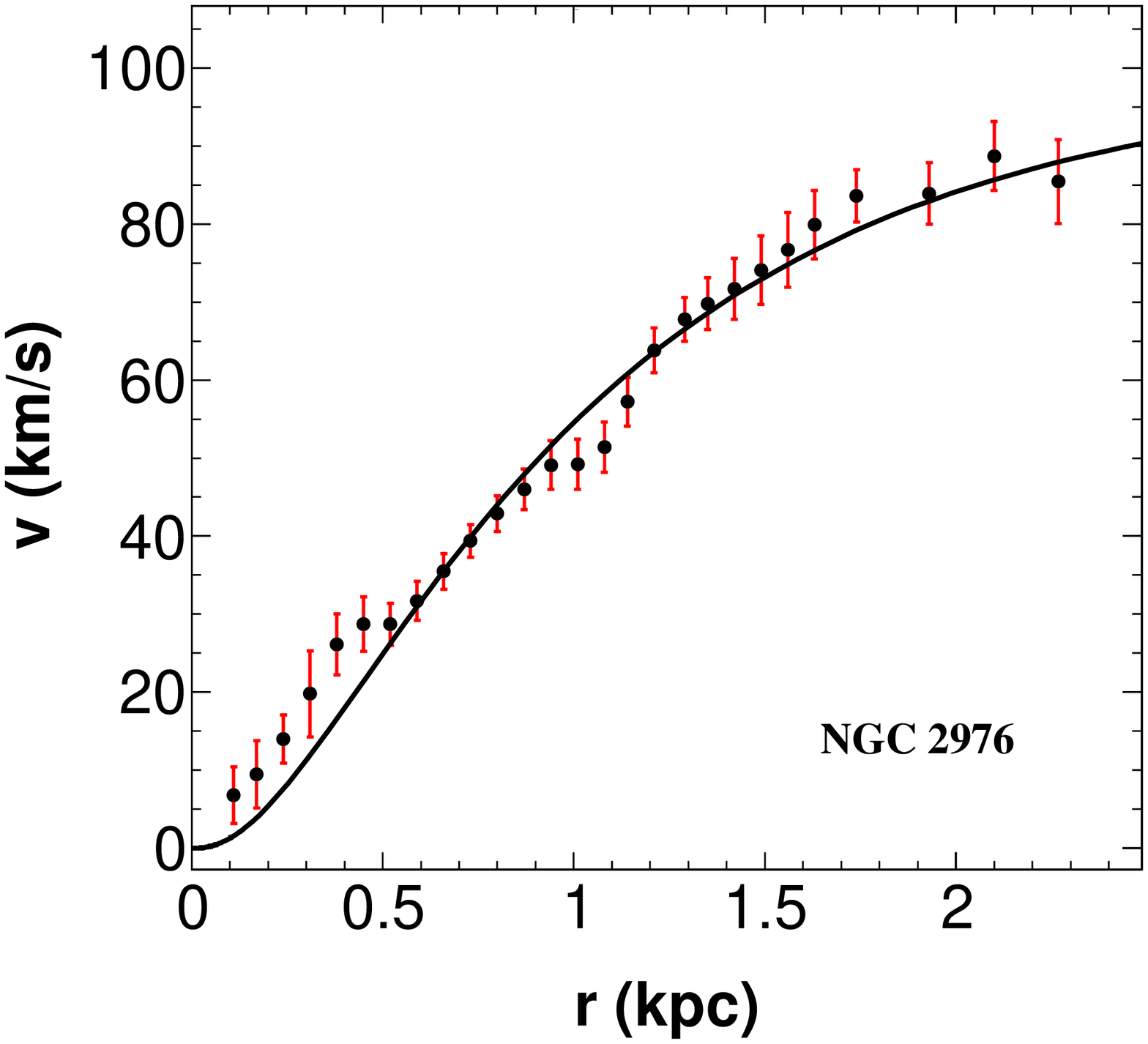} \hspace{0.2cm}
\includegraphics[scale = 0.28]{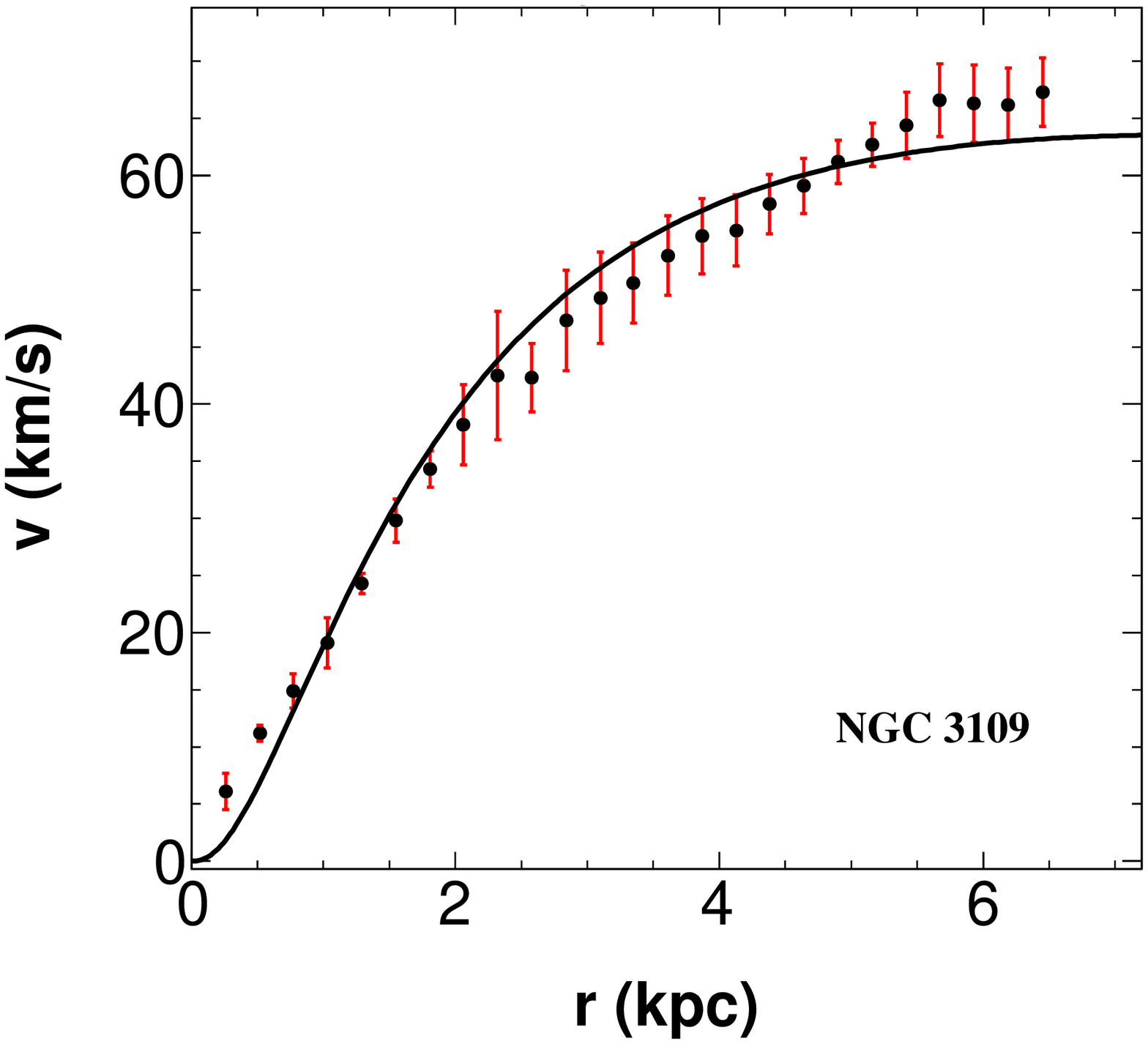} \hspace{0.2cm}
\includegraphics[scale = 0.28]{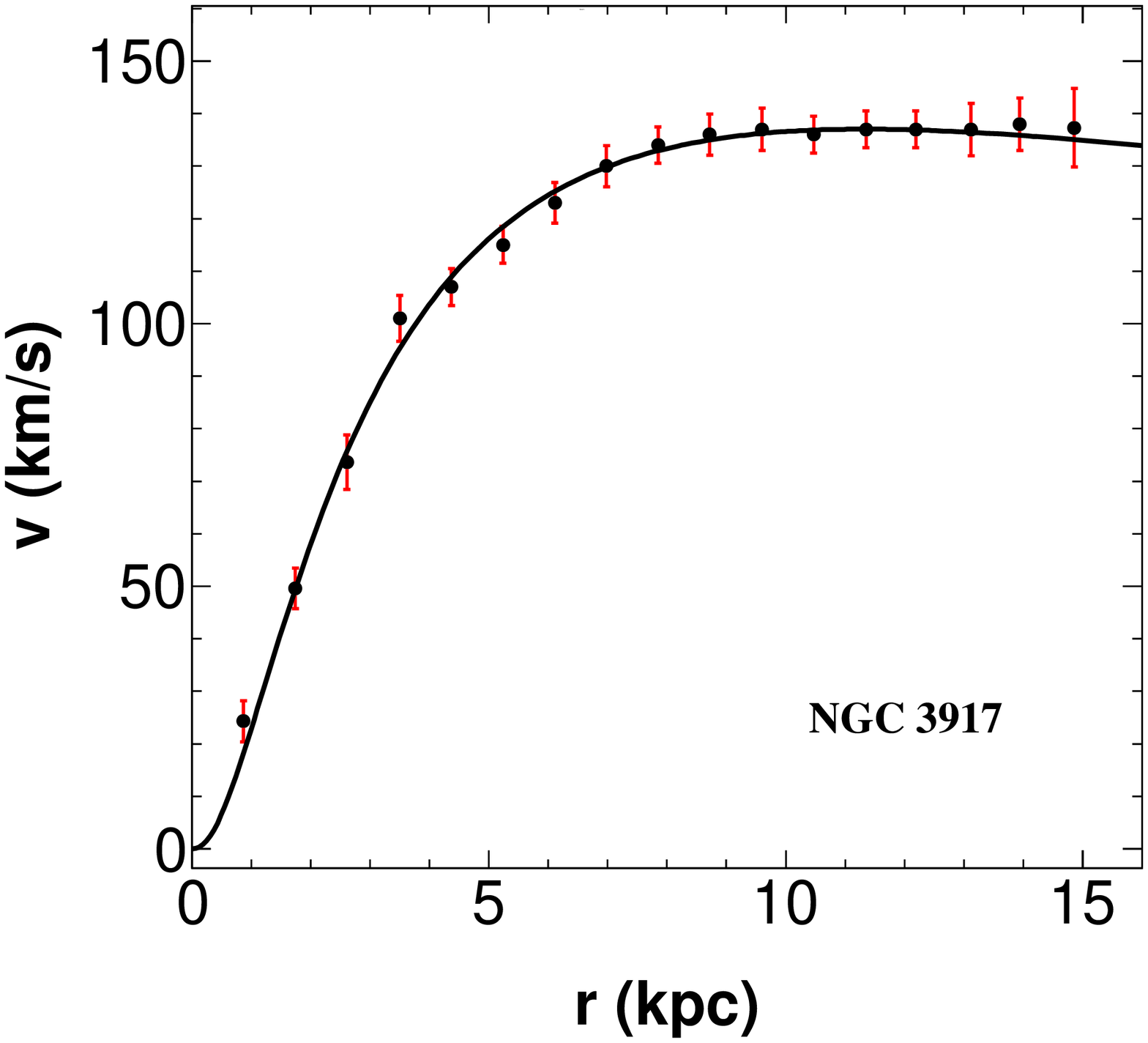} \hspace{0.2cm}
\includegraphics[scale = 0.28]{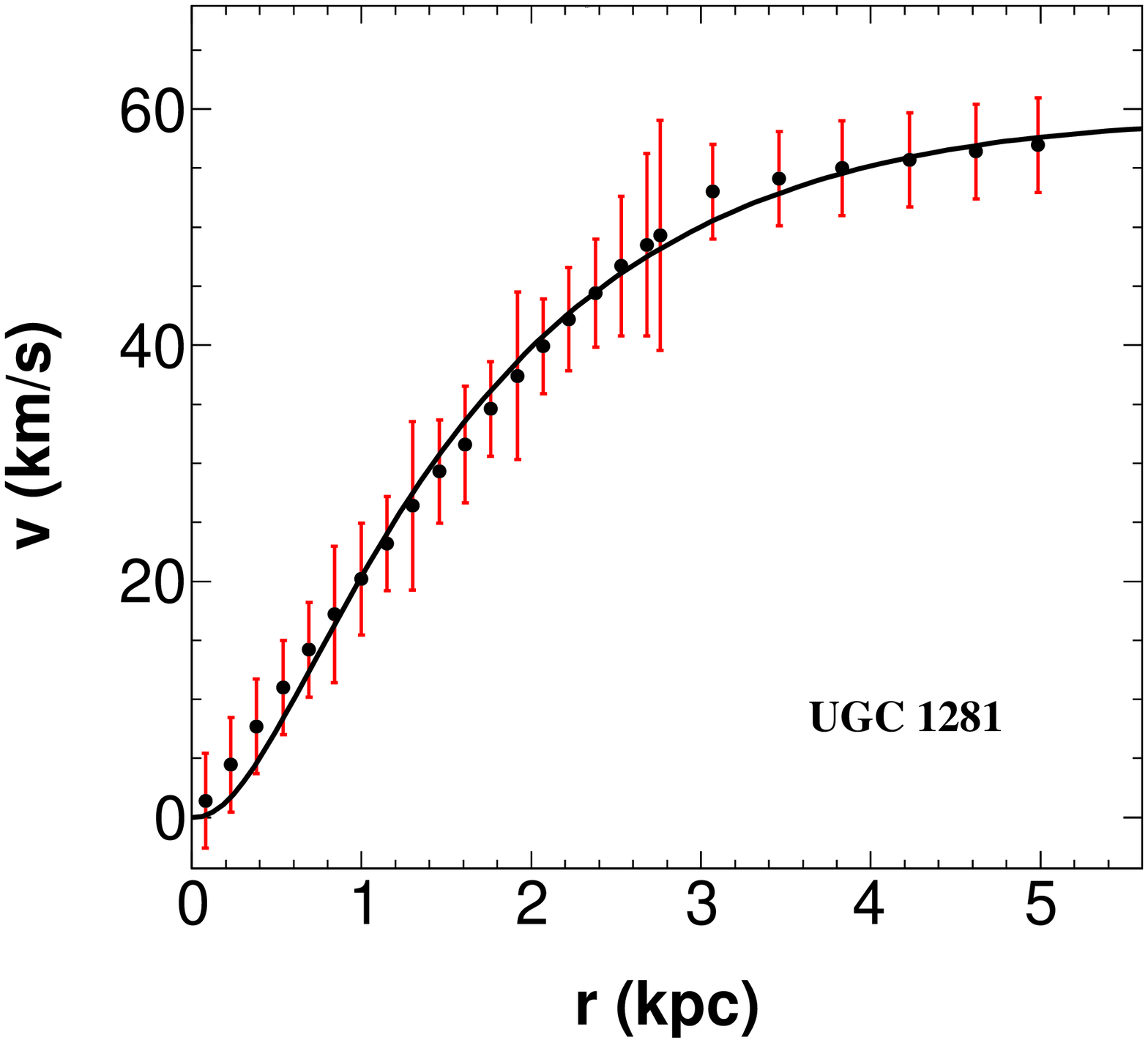} \hspace{0.2cm}
\includegraphics[scale = 0.28]{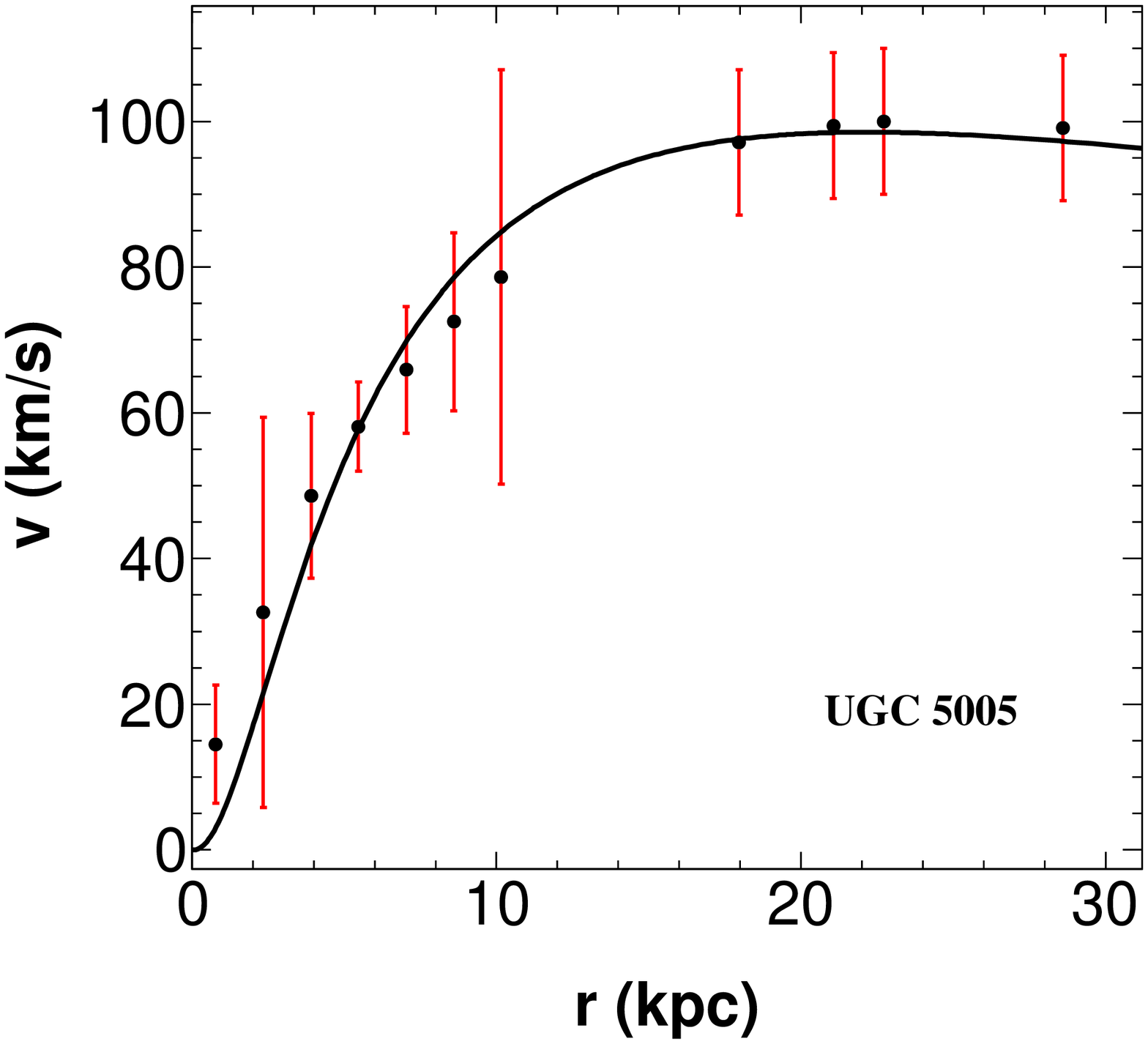} \hspace{0.2cm}
\caption{Fitting of equation \eqref{eqn.30} for the new model \eqref{eqn.27} to 
the rotational velocities (in km/s) of a set of 12 LSB galaxies extracted 
from Ref.~\cite{mannheim} with their errors plotted as a function of radial 
distance (in kpc).}
\label{fig.2}
\end{figure}

\begin{figure}[!h]
\includegraphics[scale = 0.28]{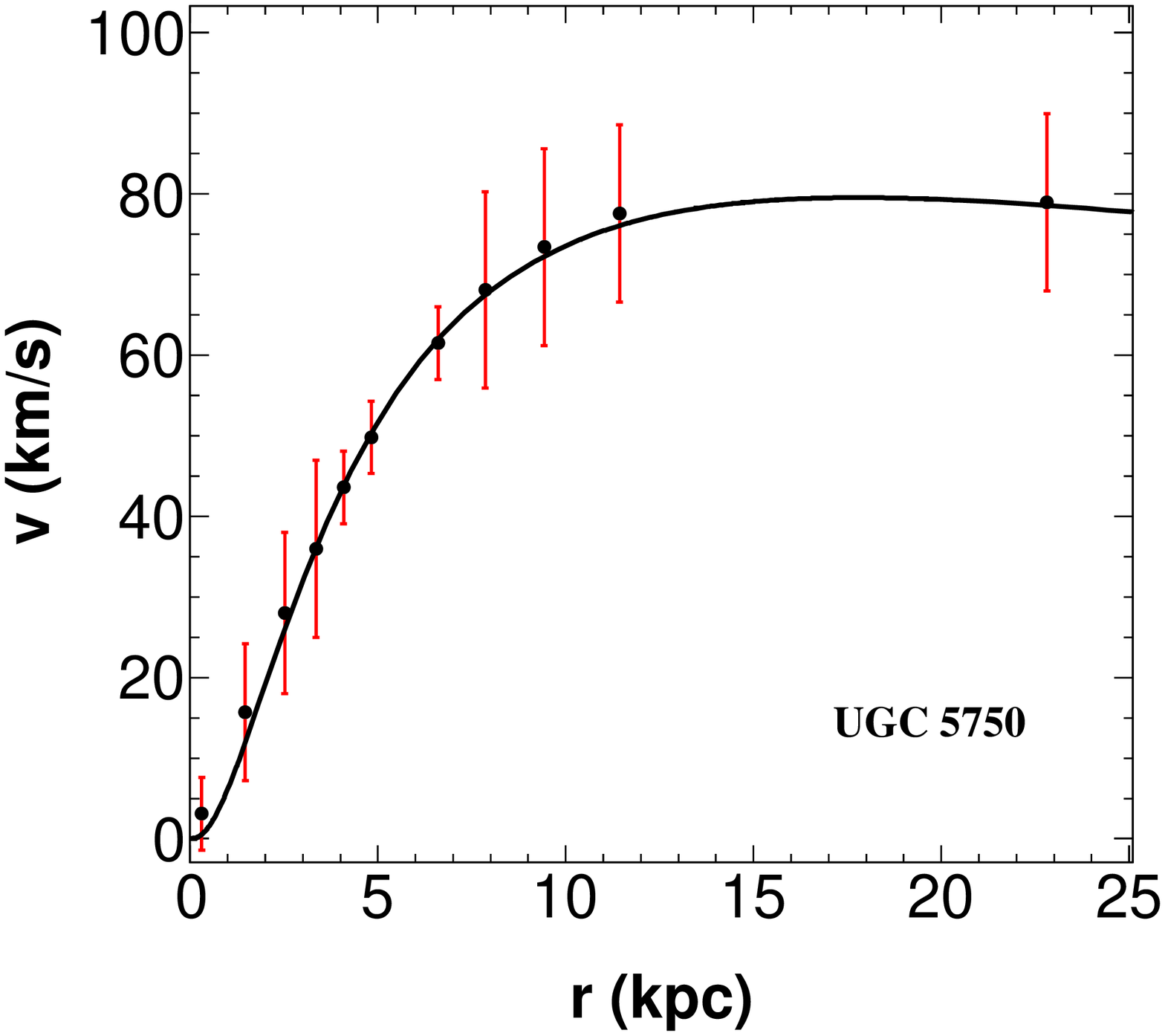} \hspace{0.2cm}
\includegraphics[scale = 0.28]{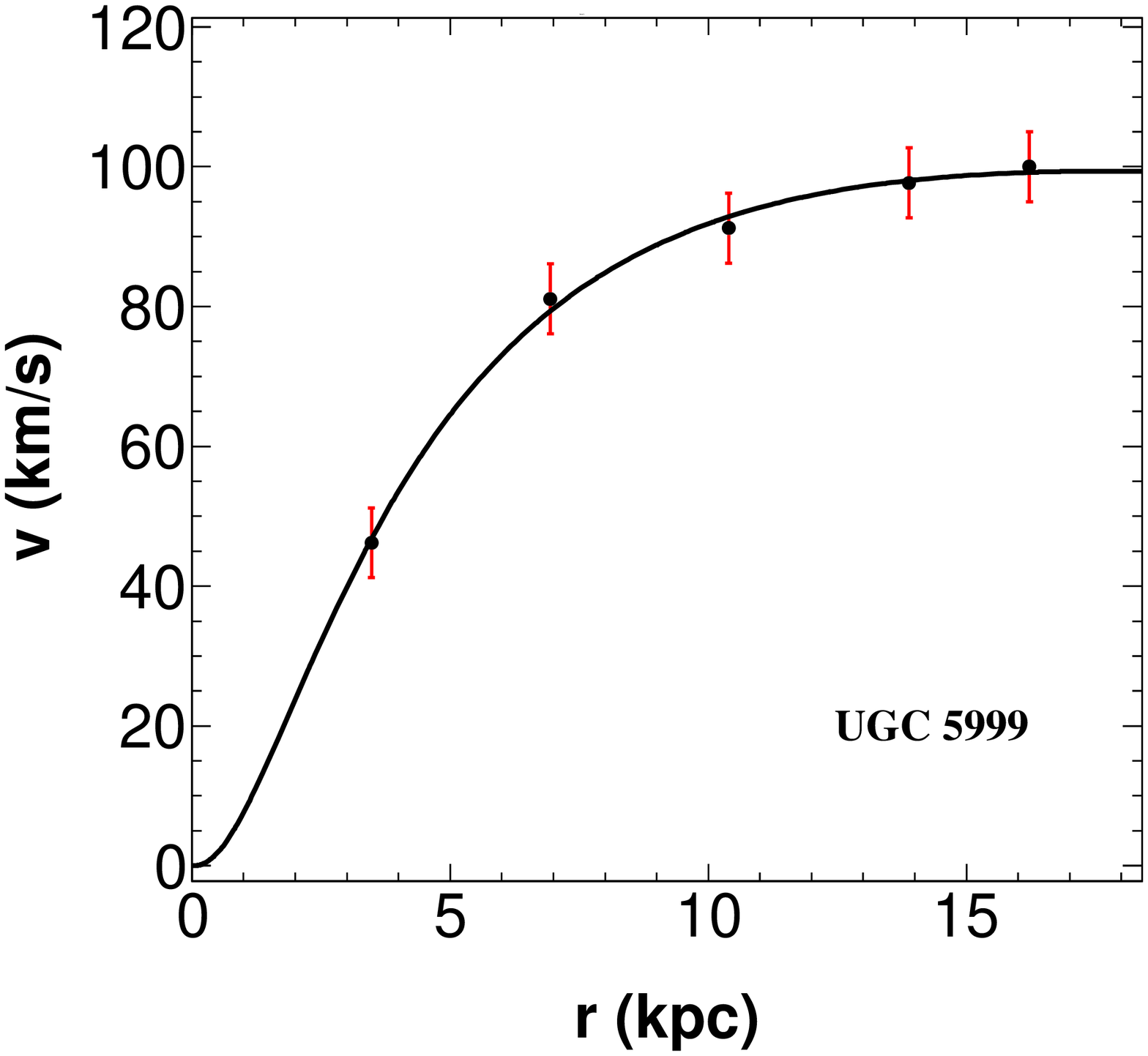} \hspace{0.2cm}
\includegraphics[scale = 0.28]{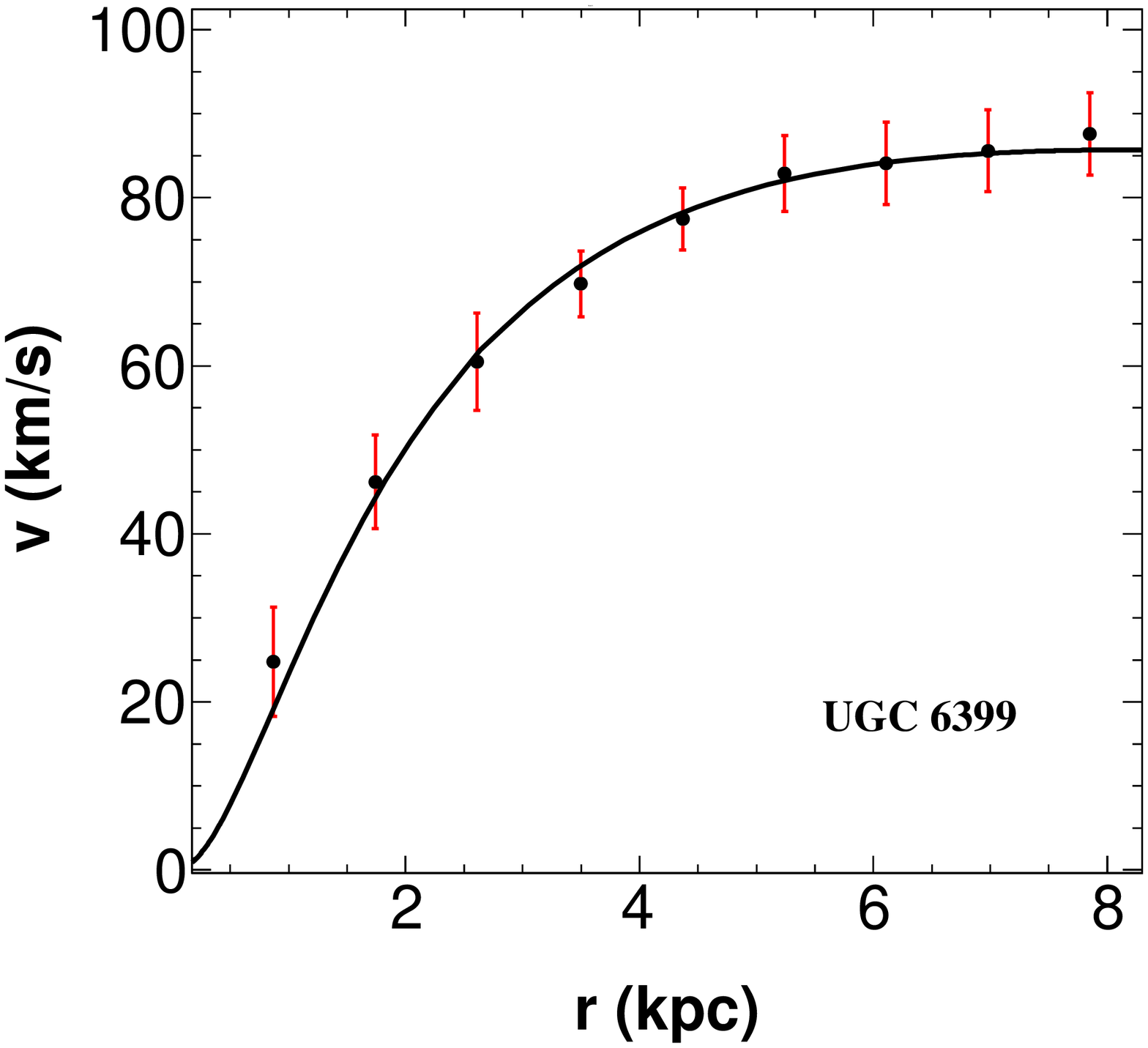} \hspace{0.2cm}
\includegraphics[scale = 0.28]{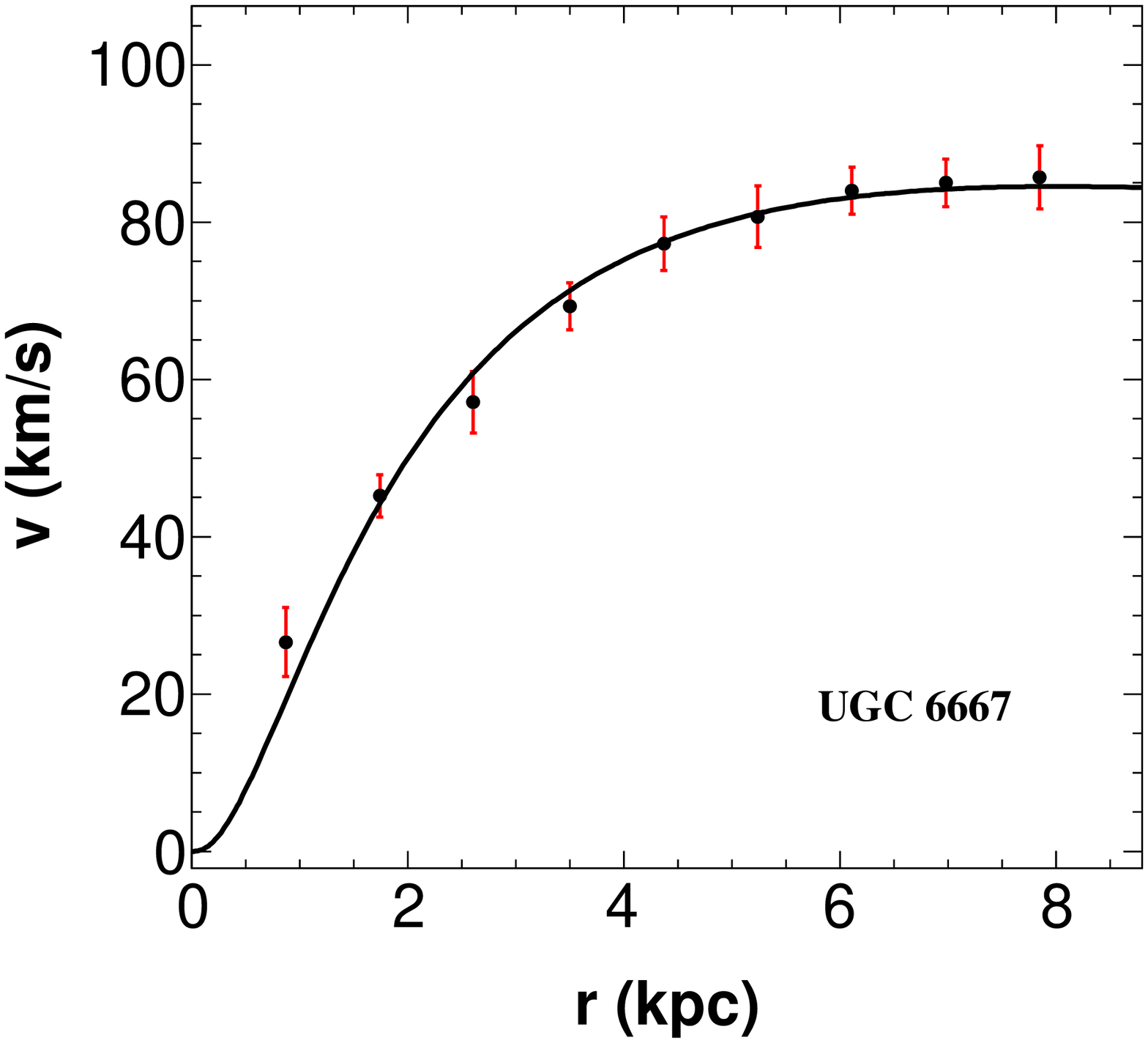} \hspace{0.2cm}
\includegraphics[scale = 0.28]{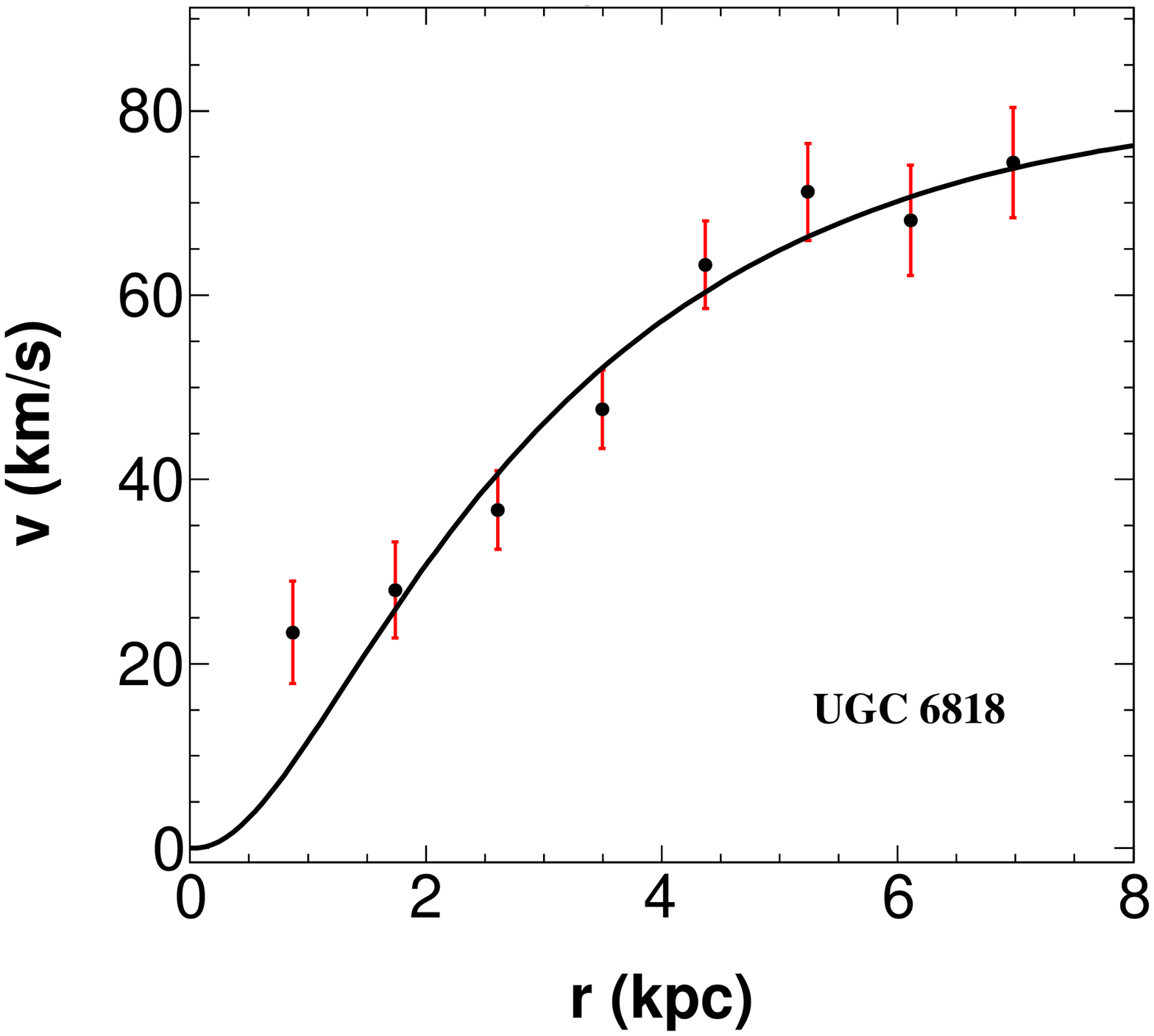} \hspace{0.2cm}
\includegraphics[scale = 0.28]{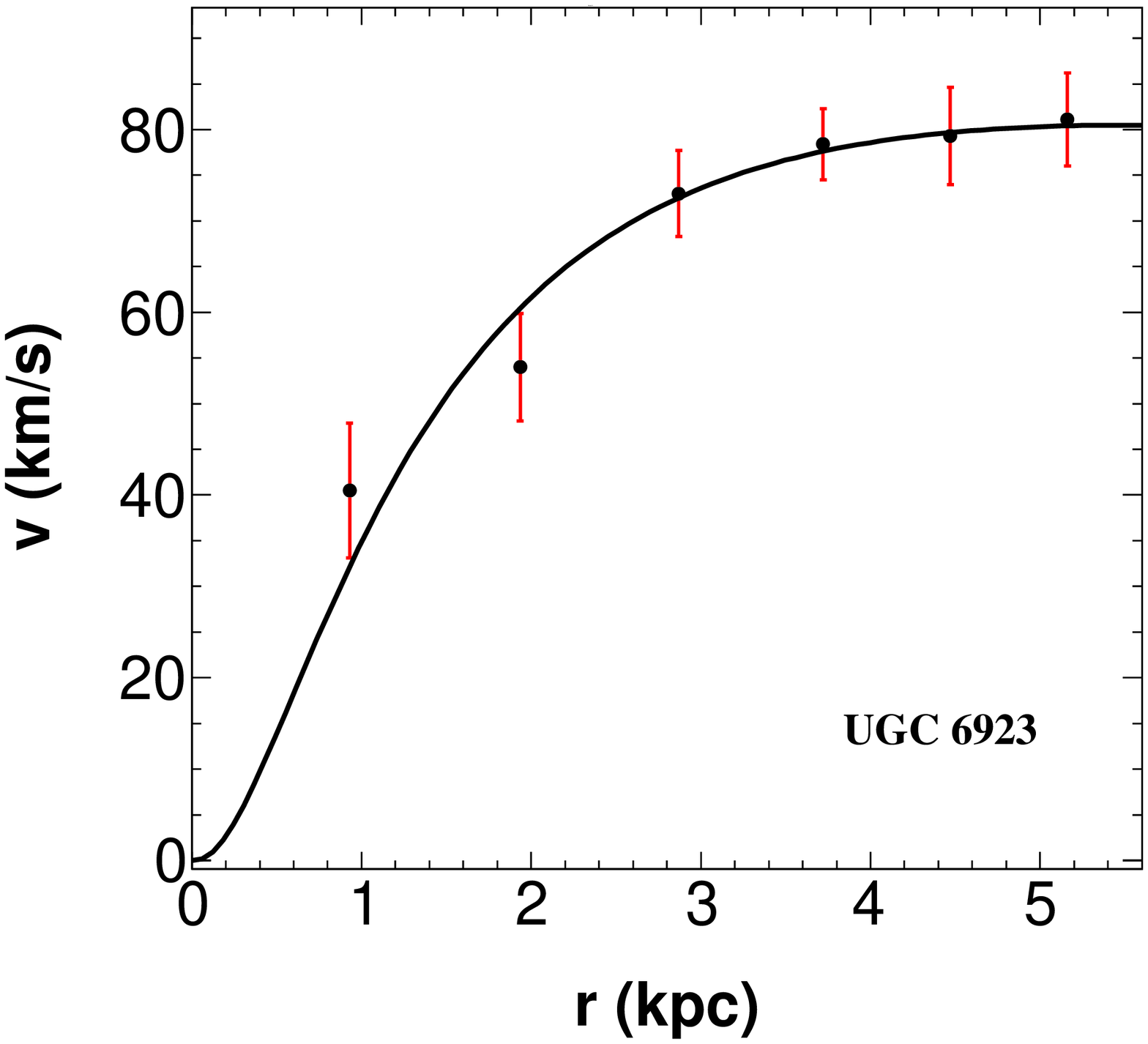} \hspace{0.2cm}
\includegraphics[scale = 0.28]{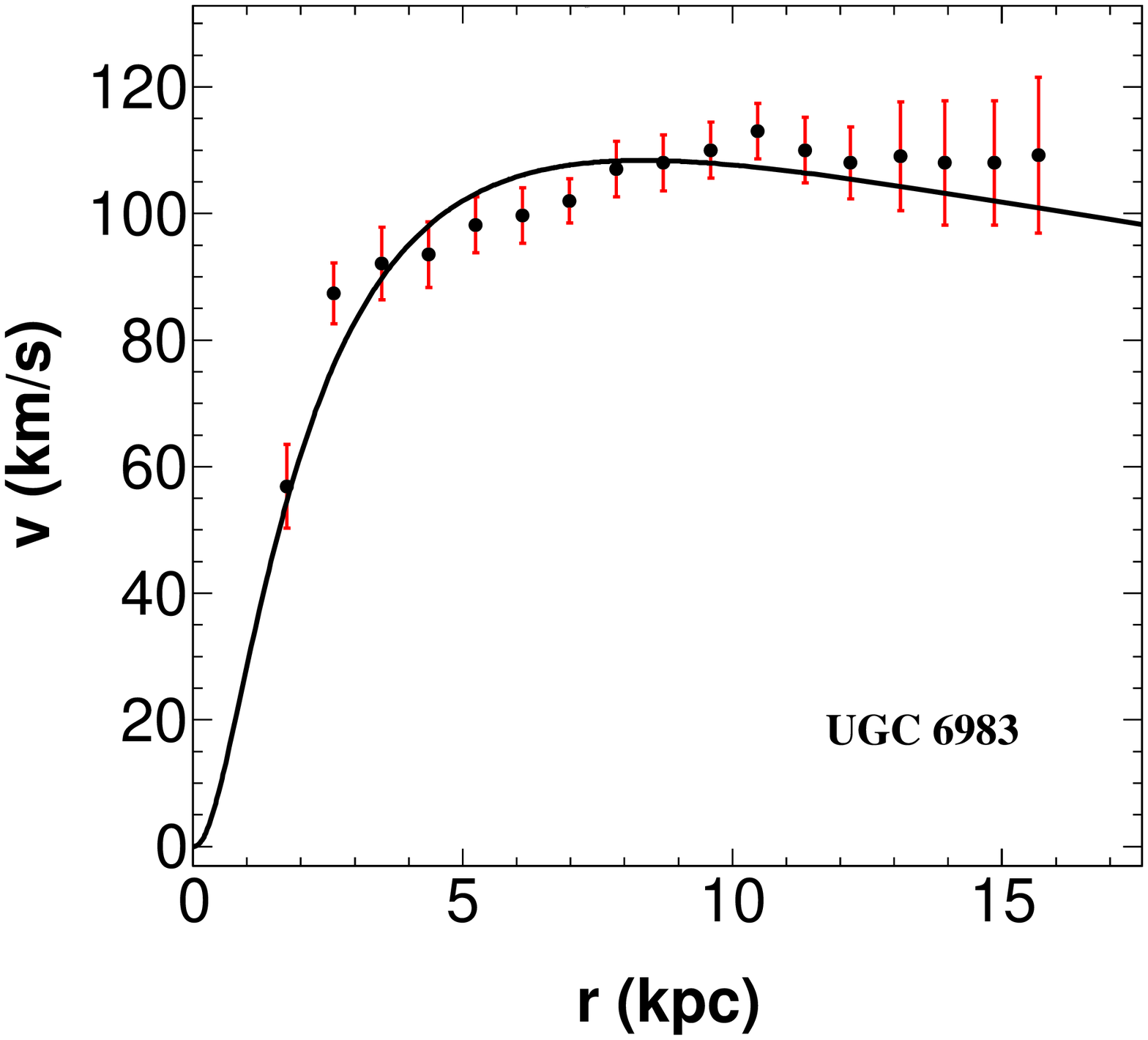} \hspace{0.2cm}
\includegraphics[scale = 0.28]{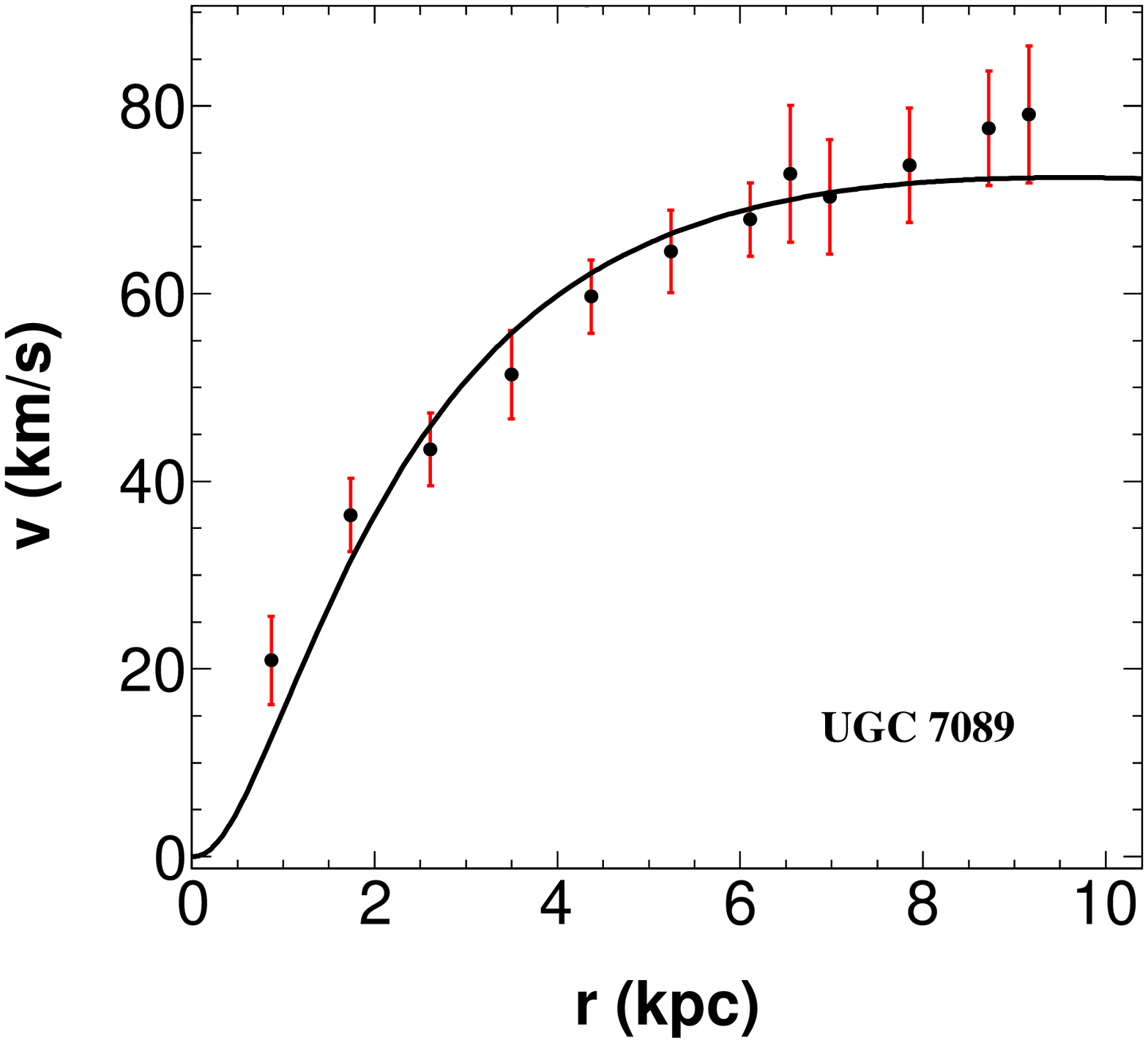} \hspace{0.2cm}
\includegraphics[scale = 0.28]{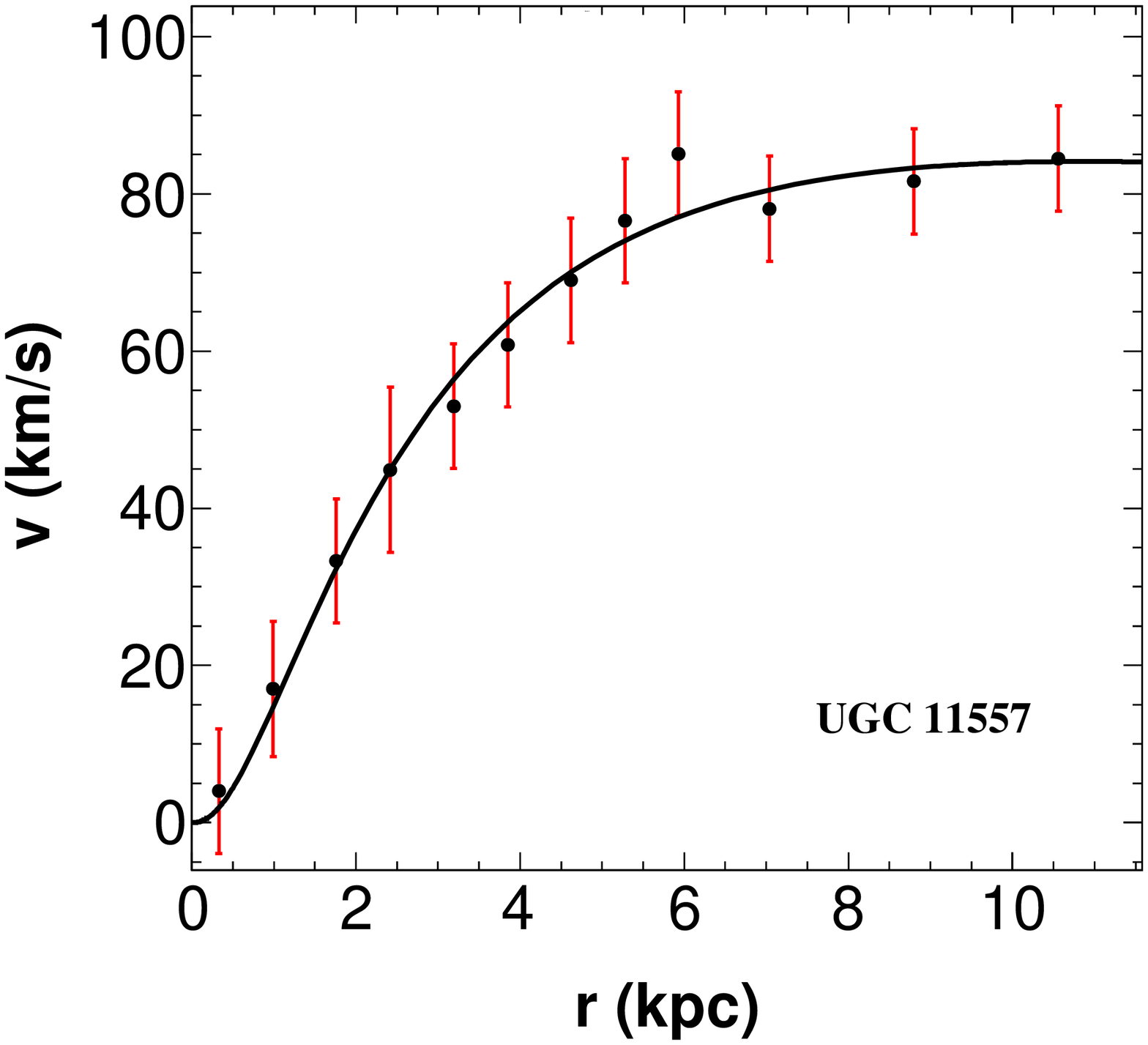} \hspace{0.2cm}
\caption{Fitting of equation \eqref{eqn.30} for the new model \eqref{eqn.27} to 
the rotational velocities (in km/s) of a set of 9 LSB galaxies extracted
from Ref.~\cite{mannheim} with their errors plotted as a function of radial 
distance (in kpc).}
\label{fig.3}
\end{figure}

\begin{table}[h!]
\begin{center}
\caption{Relevant galaxy properties for a set of nine Dwarf galaxies and the 
best fit values for the parameters $M_0$ and $r_c$ of these galaxies in the
new $f(\mathcal{R})$ gravity model \eqref{eqn.27}.}
\begin{tabular}{ c c c c c c c c c }
\hline \hline 
Galaxy & Distance & $L_B$ & $R_0 $ & $M_{HI}$ & $M_0$ & $r_c$ & $\chi_{red}^2$ & $(M/L)_\ast$ \\ \vspace{0.5mm}
  & $D$ (Mpc) & $(10^9 L_\odot)$ & (kpc) & $(10^9 M_\odot)$ & $(10^9 M_\odot)$ & (kpc) &  & $({M_\odot}/{L_\odot})$ \\ 
   \hline 
 UGC 4325 & 11.87 & 1.71 & 1.92 & 1.04 & 1.612 & 0.79 & 0.25 & 0.132\\
 UGC 4499 & 12.80 & 1.01 & 1.46 & 1.15 & 24.323 & 1.43 & 0.64 & 22.56\\
 UGC 5414 & 9.40 & 0.49 & 1.40 & 0.57 & 6.257 & 1.08 & 1.08 & 11.22\\
 UGC 7232 & 3.14 & 0.08 & 0.30 & 0.06 & 3.222 & 0.33 & 0.86 & 39.27\\
 UGC 7323 & 7.90 & 2.39 & 2.13 & 0.70 & 16.275 & 1.59 & 2.75 & 6.41\\
 UGC 7524 & 4.12 & 1.37 & 3.02 & 1.34 & 9.921 & 1.90 & 2.13 & 5.94\\
 UGC 7559 & 4.20 & 0.04 & 0.87 & 0.12 & 1.279 & 0.69 & 0.16 & 27.99\\
 UGC 7577 & 3.03 & 0.10 & 0.73 & 0.06 & 0.895 & 0.73 & 0.17 & 8.15\\
 UGC 12632 & 9.20 & 0.86 & 3.43 & 1.55 & 3.487 & 1.68 & 0.75 & 1.65\\
 \hline \hline   
\end{tabular}
\label{Table:3}
\end{center}
\end{table}

\subsection{Analysis of LSB galaxy sample}

It is well known that LSB galaxies are regarded as dark matter dominated 
galaxies and hence, they can provide a satisfactory test for a gravitational 
theory \cite{deblok2002, mannheim}. To this end, we fit equation 
\eqref{eqn.30} to a sample of 21 LSB galaxies extracted from 
Ref.~\cite{mannheim} with their maximum radial distance varying from $2.1$ kpc 
to $18.2$ kpc. Table \ref{Table:2} compiles the relevant data for these 
galaxies we have selected from the $111$ galaxies studied in 
Ref.~\cite{mannheim} along with the respective best fit values of the 
parameters $M_0$ and $r_c$ and the $\chi_{red}^2$ values of the fits. The 
results are presented in Fig.\ \ref{fig.2} and \ref{fig.3}. 

It is seen from Table \ref{Table:2} that the mass-to-light ratios for few 
galaxy samples are much larger than the upper bound 
$10 ({M_\odot}/{L_\odot})$ \cite{sanders}. However, as can be seen from 
Figs.\ \ref{fig.2} and \ref{fig.3}, the rotation curves are well fitted with 
the observed data as the $\chi_{red}^2$ values are smaller than or equivalent 
to $1$ for almost all of the galaxies in the sample. For the galaxies NGC 
$0300$, NGC $3109$ and UGC $6818$ although the $\chi_{red}^2$ values are high, 
around $2.04$, $2.78$ and $1.69$ respectively, these do not affect the shape 
of the fitted rotation curves.

\begin{figure}[!h]
\includegraphics[scale = 0.28]{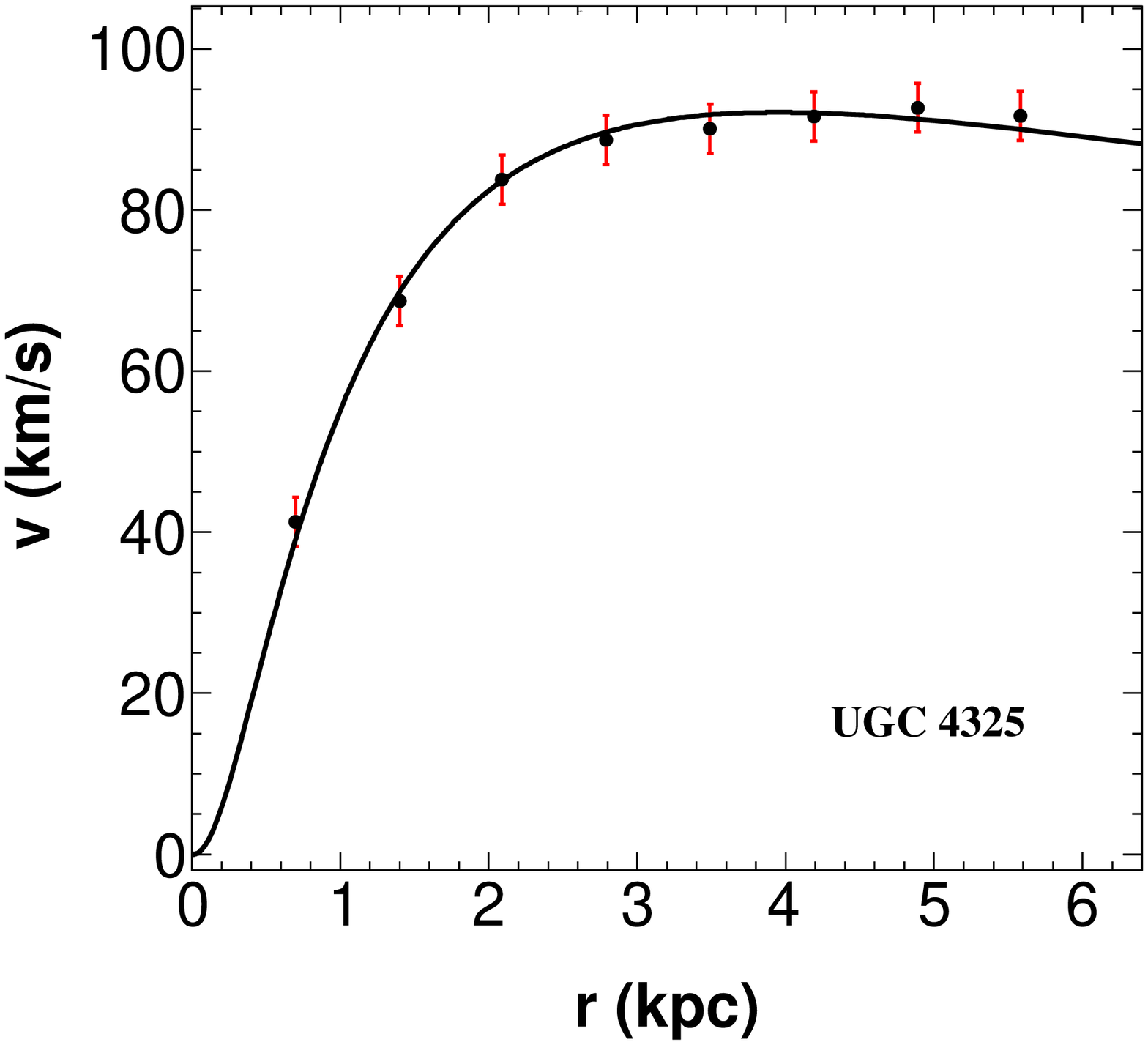} \hspace{0.2cm}
\includegraphics[scale = 0.28]{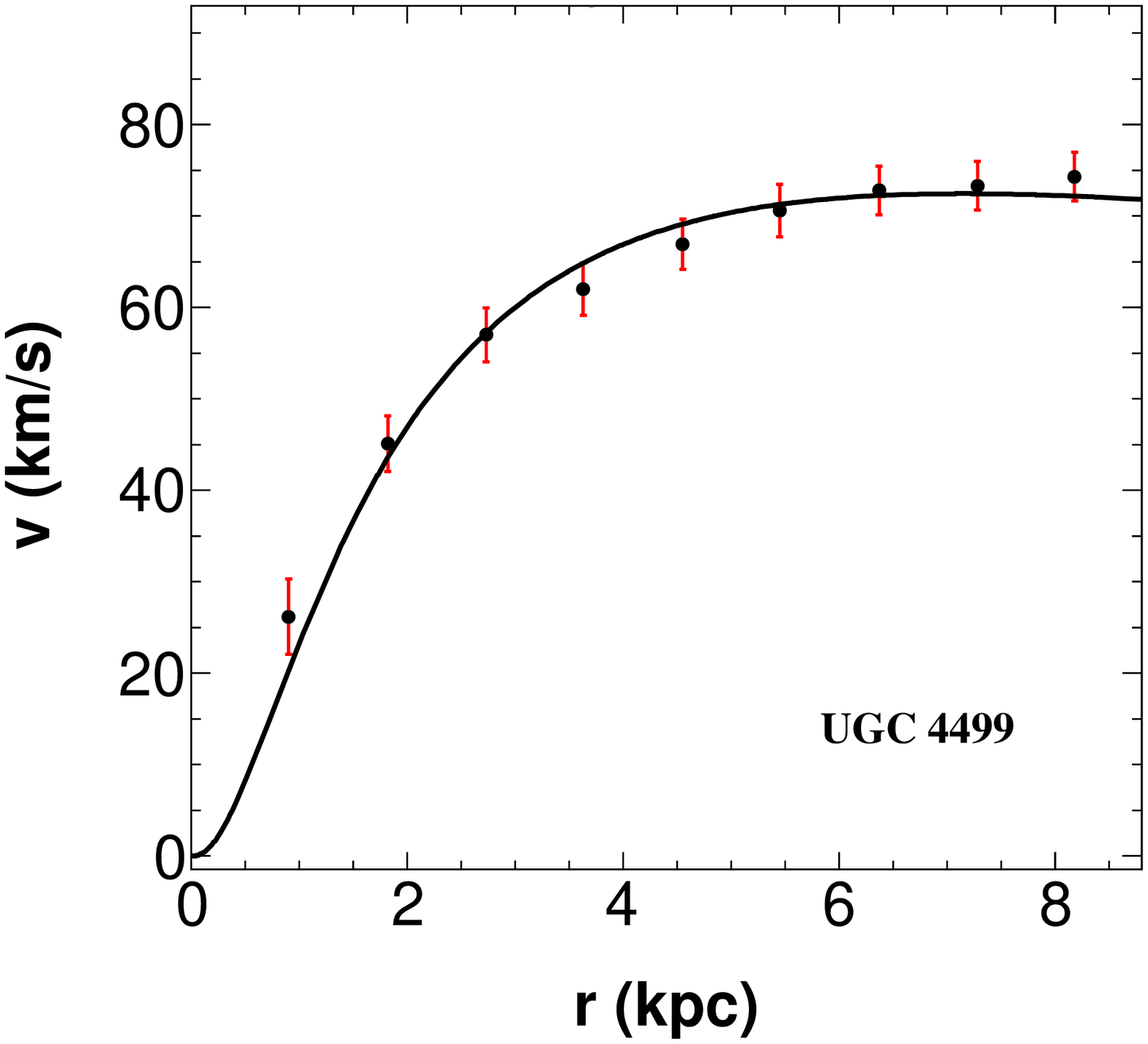} \hspace{0.2cm}
\includegraphics[scale = 0.28]{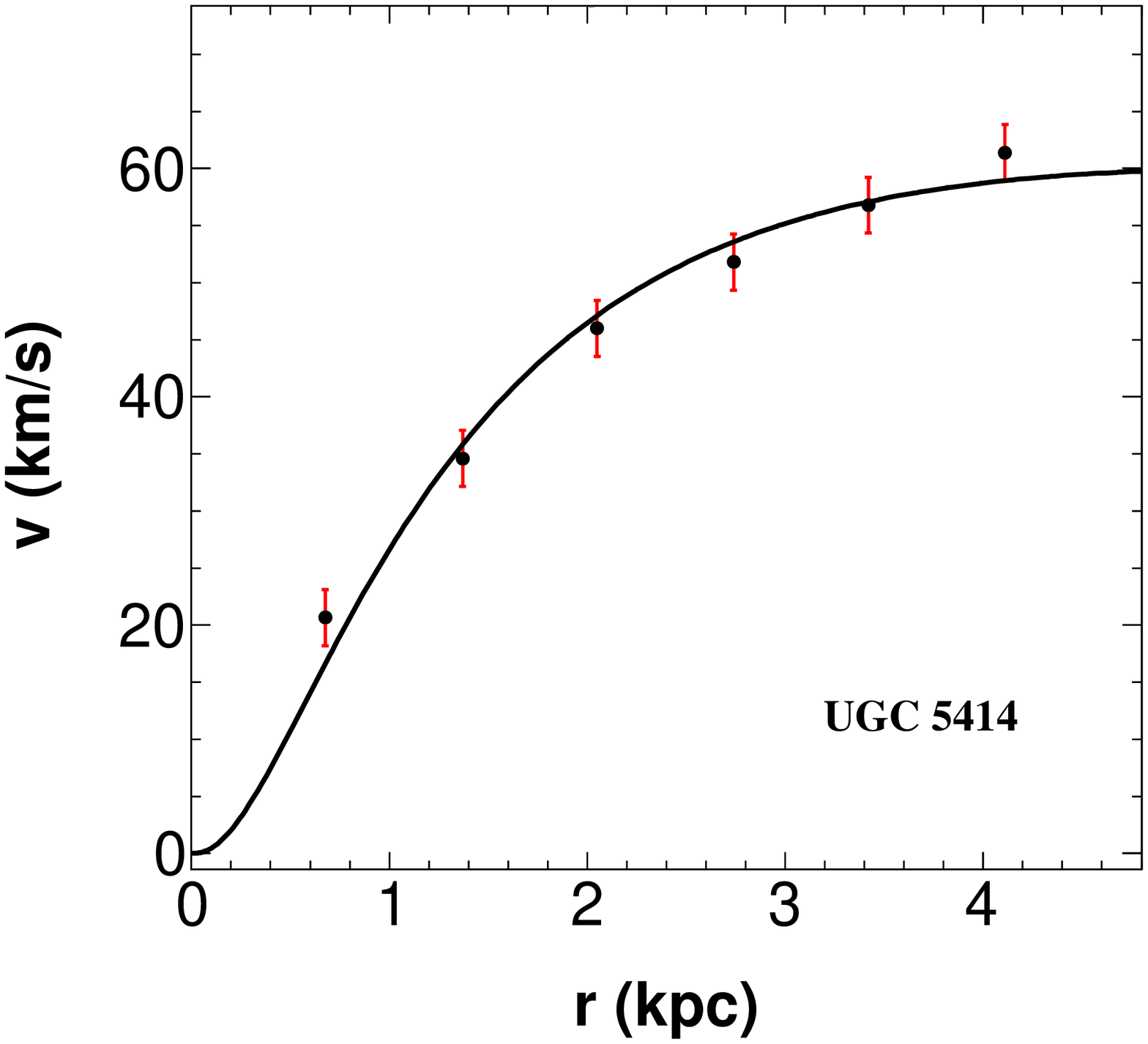} \hspace{0.2cm}
\includegraphics[scale = 0.28]{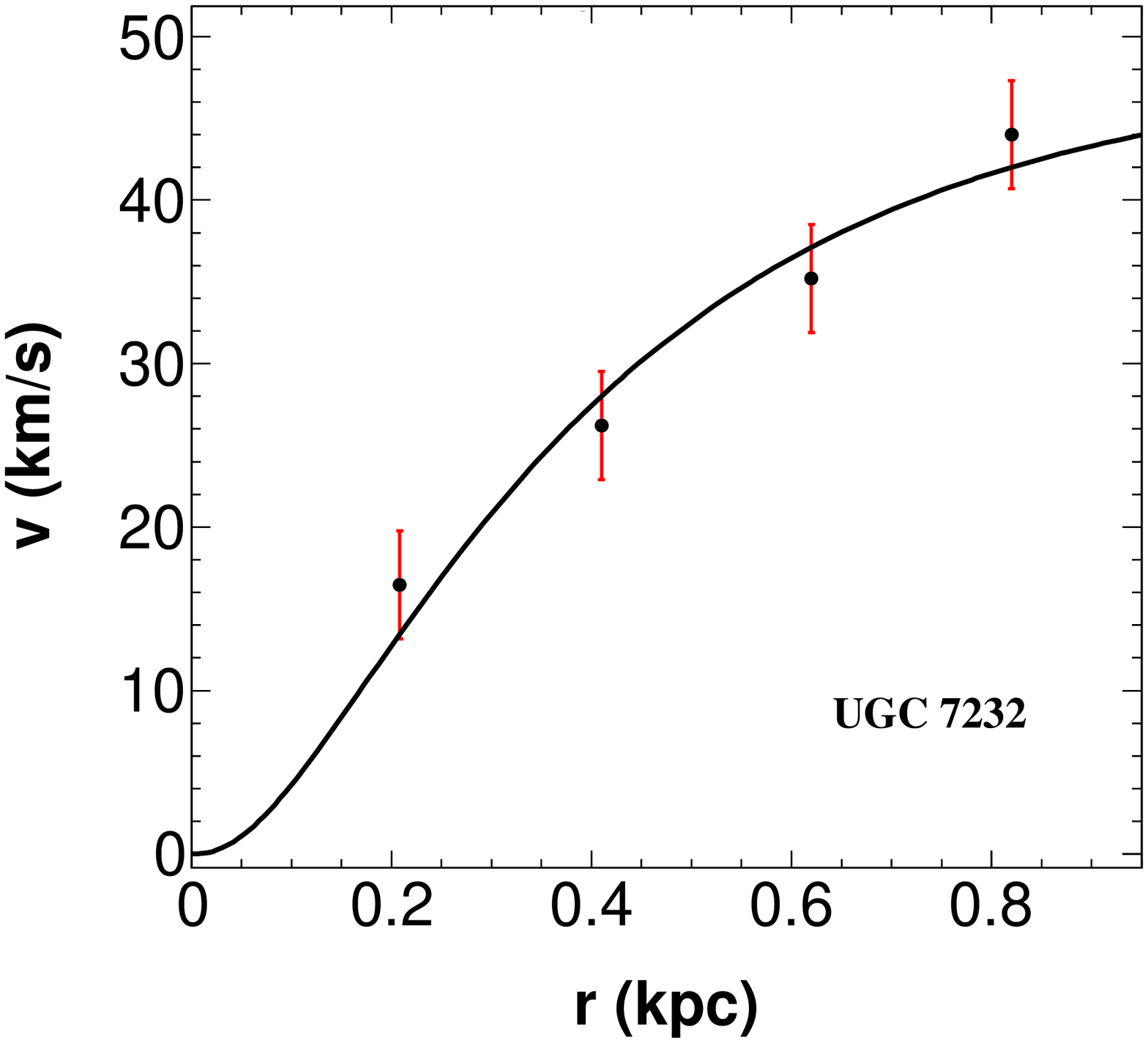} \hspace{0.2cm}
\includegraphics[scale = 0.28]{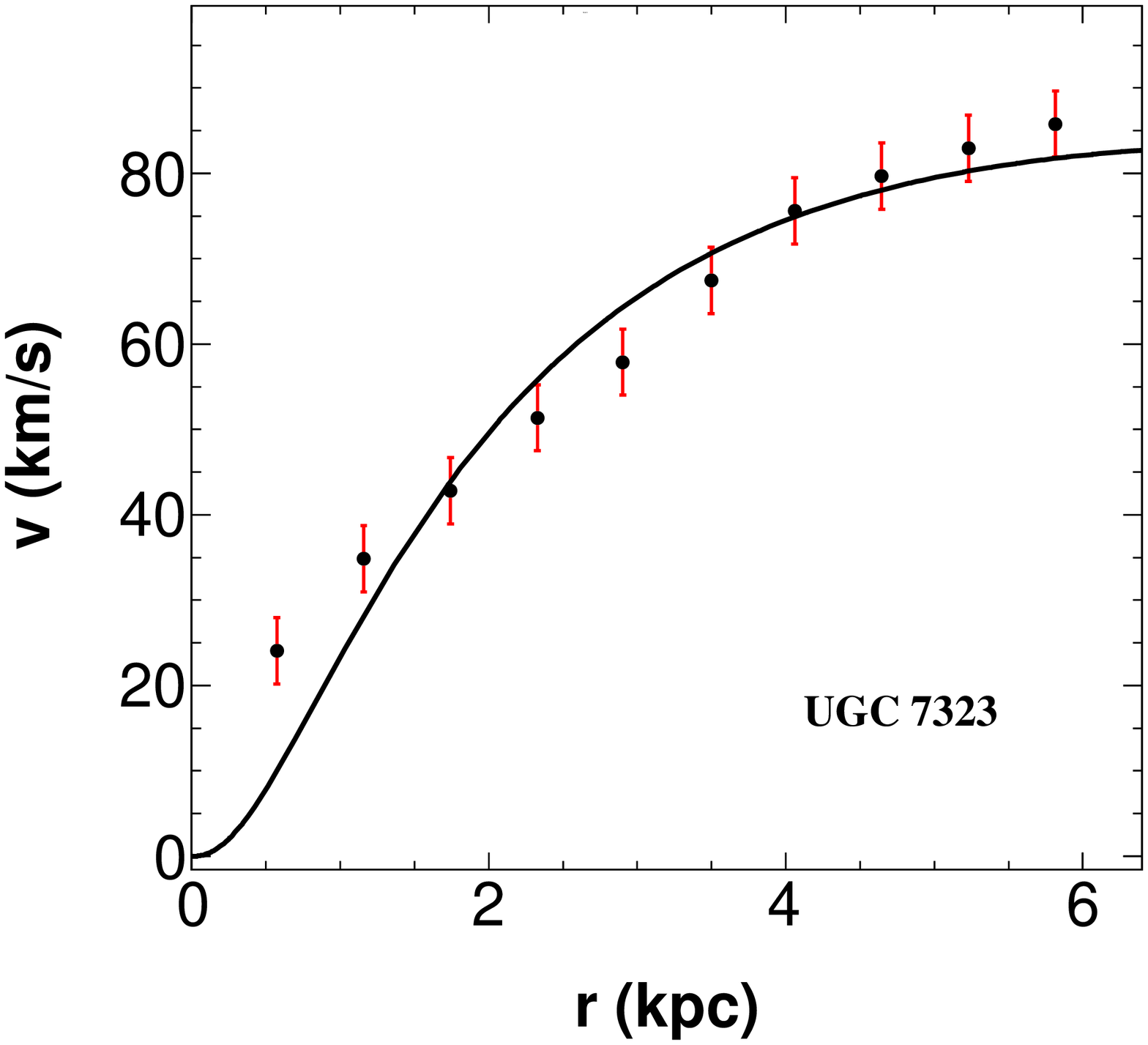} \hspace{0.2cm}
\includegraphics[scale = 0.28]{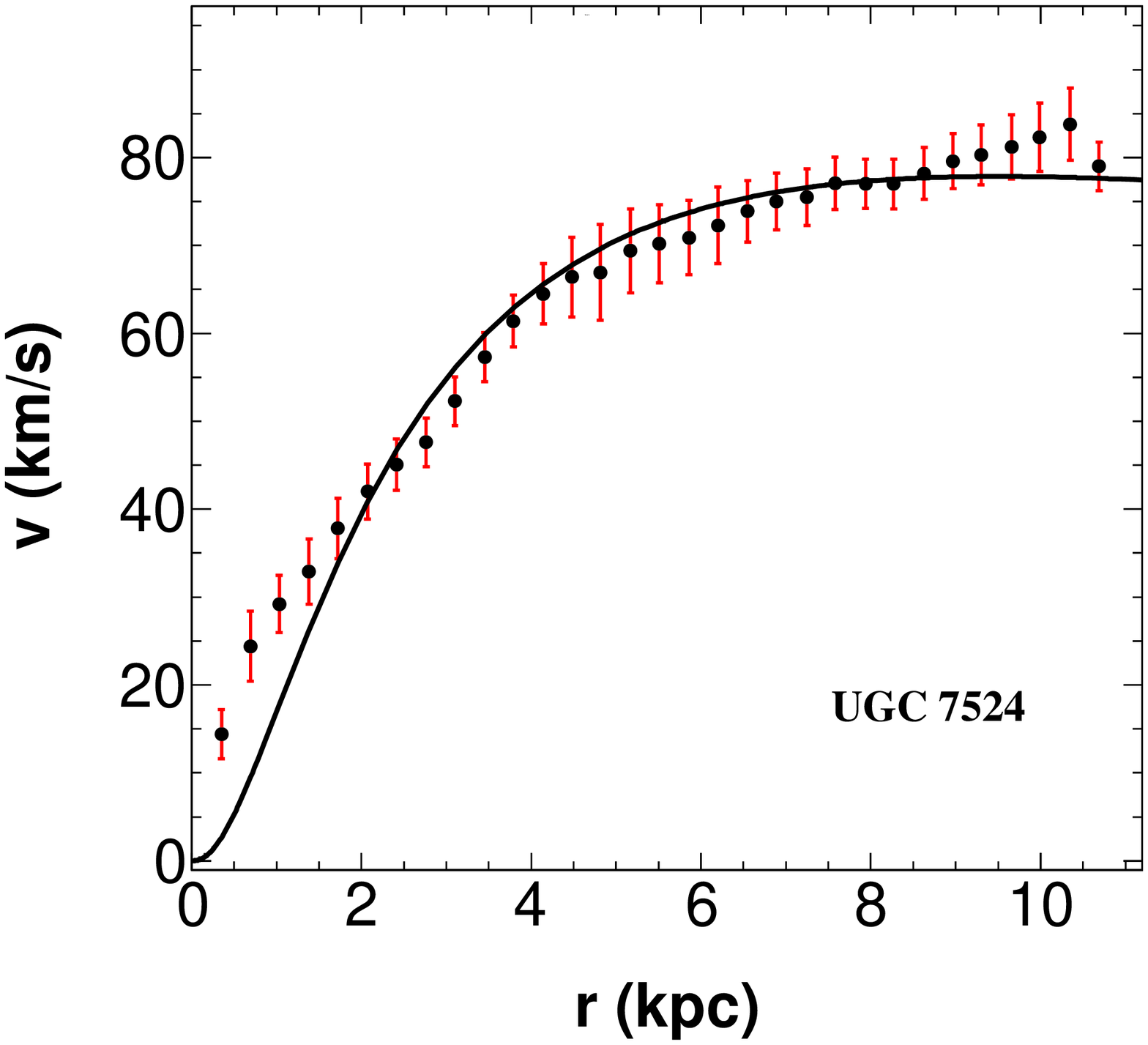} \hspace{0.2cm}
\includegraphics[scale = 0.28]{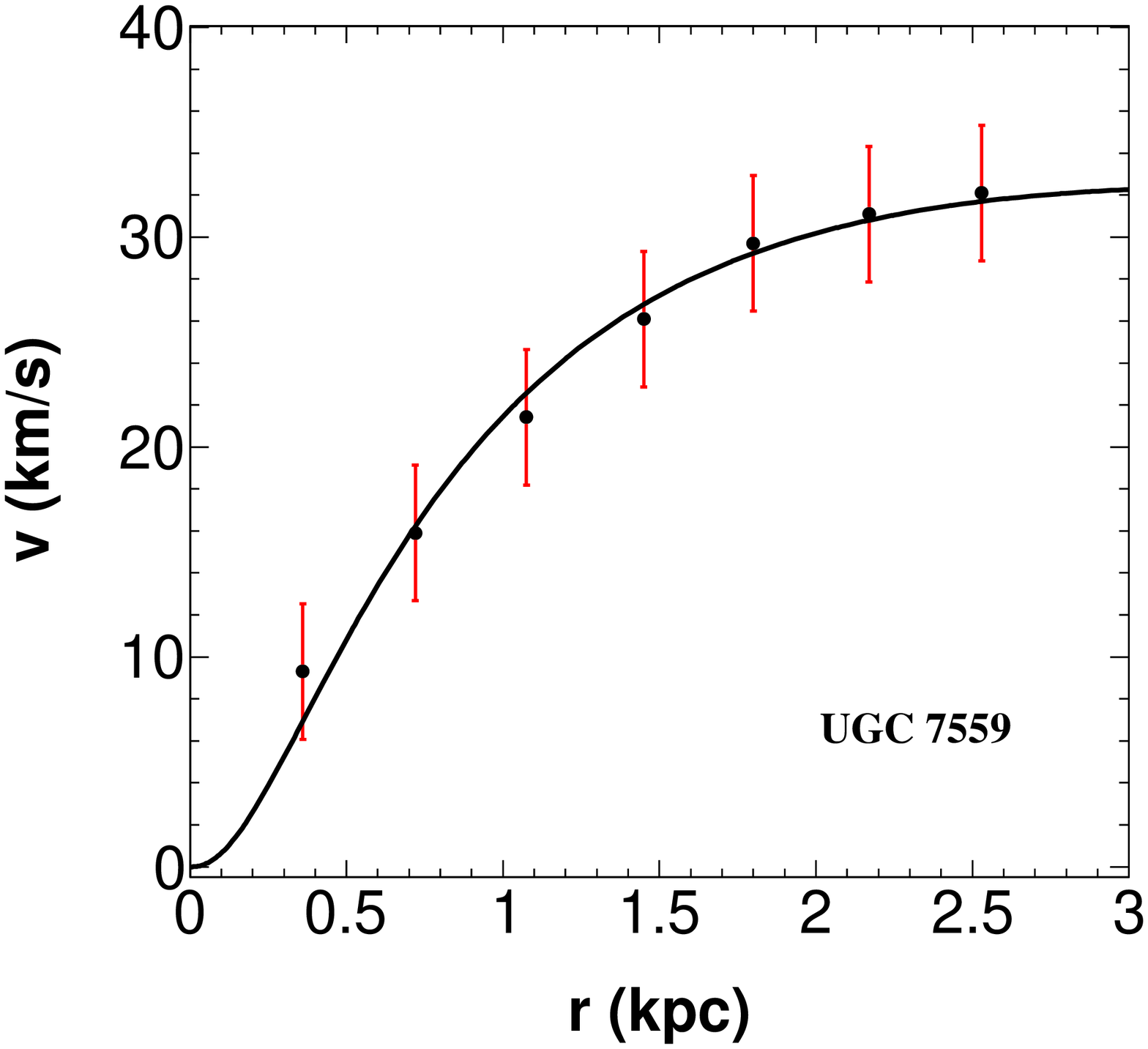} \hspace{0.2cm}
\includegraphics[scale = 0.28]{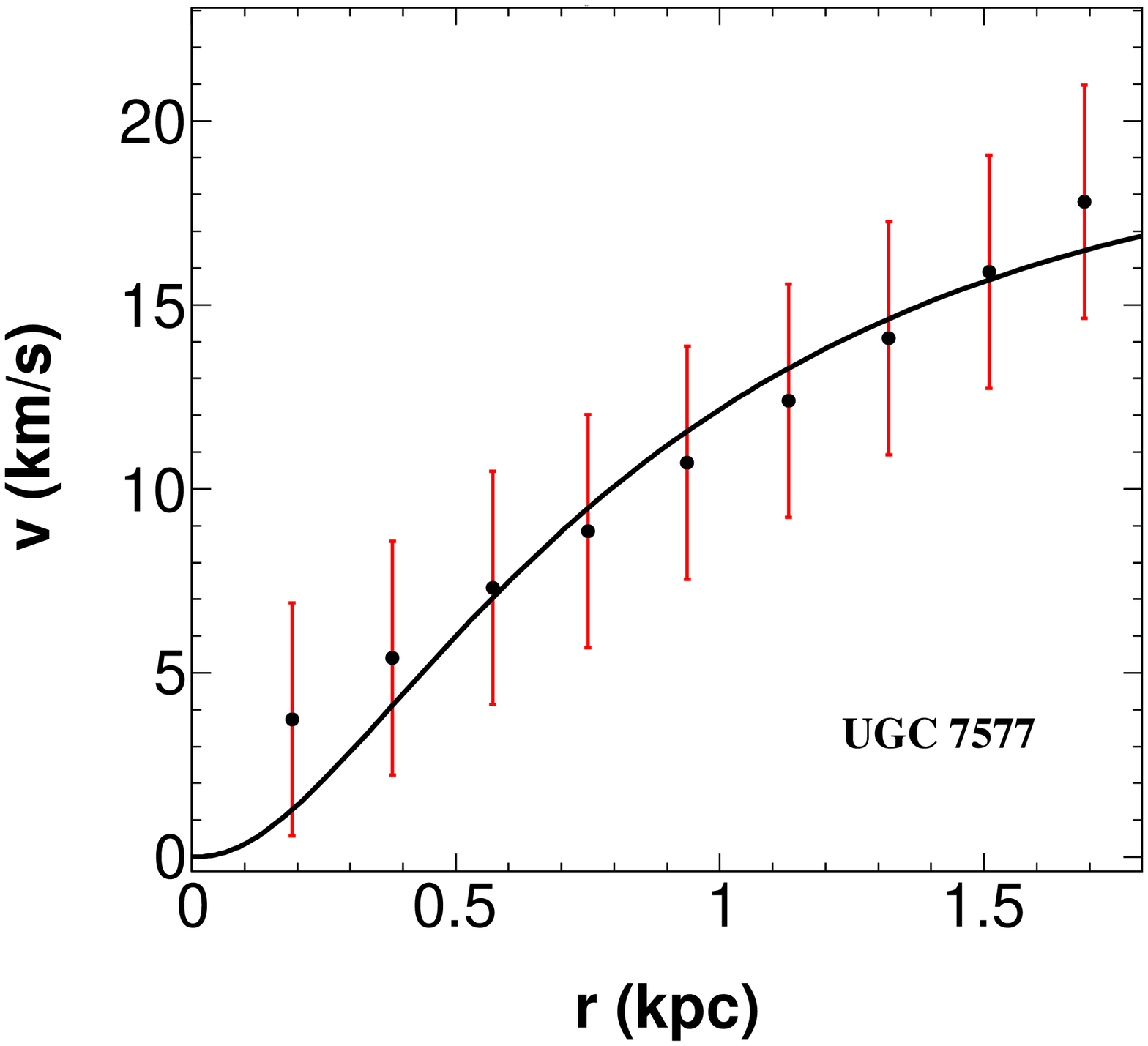} \hspace{0.2cm}
\includegraphics[scale = 0.28]{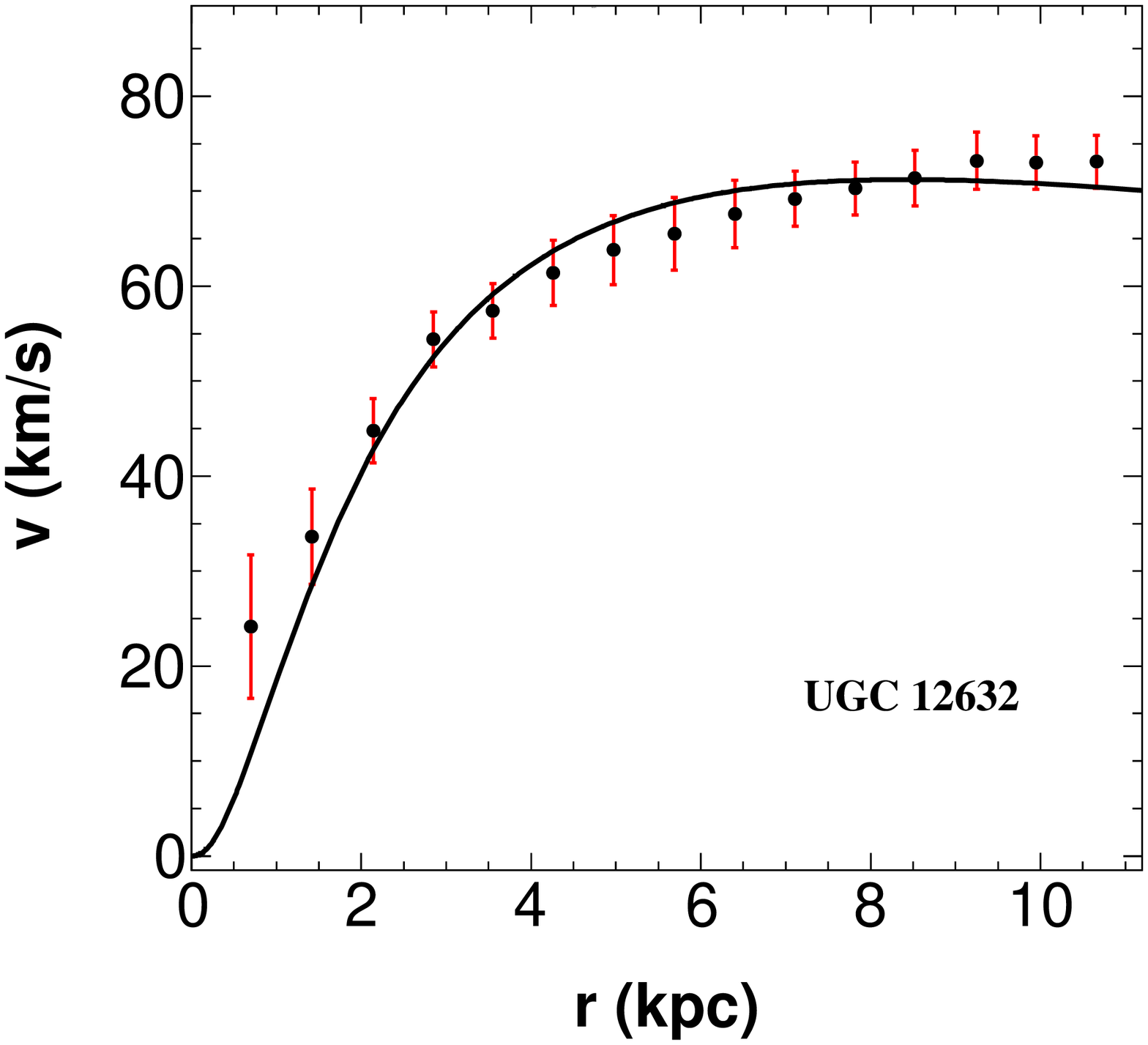} \hspace{0.2cm}
\caption{Fitting of equation \eqref{eqn.30} for the new model \eqref{eqn.27} to the 
rotational velocities (in km/s) of a set of nine dwarf galaxies extracted 
from Ref.~\cite{mannheim} with their errors plotted as a function of radial 
distance (in kpc).}
\label{fig.4}
\end{figure}

\subsection{Analysis of dwarf galaxy sample}

For dwarf galaxies, we have chosen a sample of nine galaxies taken from 
Ref.~\cite{obrien}. Table \ref{Table:3} compiles the relevant data for the 
chosen galaxy sample and the results are depicted in Fig.\ \ref{fig.3}. For 
all the galaxies the $\chi_{red}^2$ values are small and gives well fitted 
galactic rotation curves, except for two galaxies viz., UGC $7323$ and UGC 
$7524$. Furthermore, the mass-to-light ratio for most of the galaxies are 
within the upper bound. However, for two galaxies namely, UGC $4499$ and 
UGC $7559$, the mass-to-light ratios are notably greater than the upper bound 
value as expected from the population synthesis models \cite{sanders}. 
According to the population synthesis models depending on the history of the 
star formation and also on metallicities the blue-band mass-to-light ratio may
range from a few tenths to 10. 

\subsection{Analysis of an ultra diffuse galaxy}

Here, we analyze the rotation curve of an ultra diffuse galaxy (UDG), AGC 
$242019$, which was identified by the Arecibo Legacy Fast ALFA (ALFALFA) 
survey of HI galaxies \cite{leisman}. This galaxy has an HI mass of 
$8.51 \times 10^8$ M$_\odot$ and a corresponding distance of $30.8$ Mpc. 
The gas rich UDG AGC $242019$ has been claimed to be like an observed LSB 
galaxy with a slowly rising rotation curve \cite{shi21}. Fig.\ \ref{fig.4} 
shows the fitted rotation curve with the data taken from Ref.~\cite{shi21}. It 
can be seen that the curve of this galaxy is well fitted with observation. In 
fact, from Fig.\ \ref{fig.4} one can see that the rotation curve is slowly 
rising, indicating that UDG AGC $242019$ can be considered as a member of a 
class of LSB galaxies. However, further study in this regard is still required 
to confirm such proclamation. From our study, the $\chi_{red}^2$ value is 
obtained as $6.84$ and the fitted parameters are evaluated to be 
$M_0 = 4.92$ ($10^{10} M_\odot$) and $r_c = 2.29$ (kpc). Thus, the flat 
rotation curve of UDG AGC $242019$ can be explained via the new 
model \eqref{eqn.27} of $f(\mathcal{R})$ gravity. 

\begin{figure}[h]
    \centering
    \includegraphics[scale = 0.38]{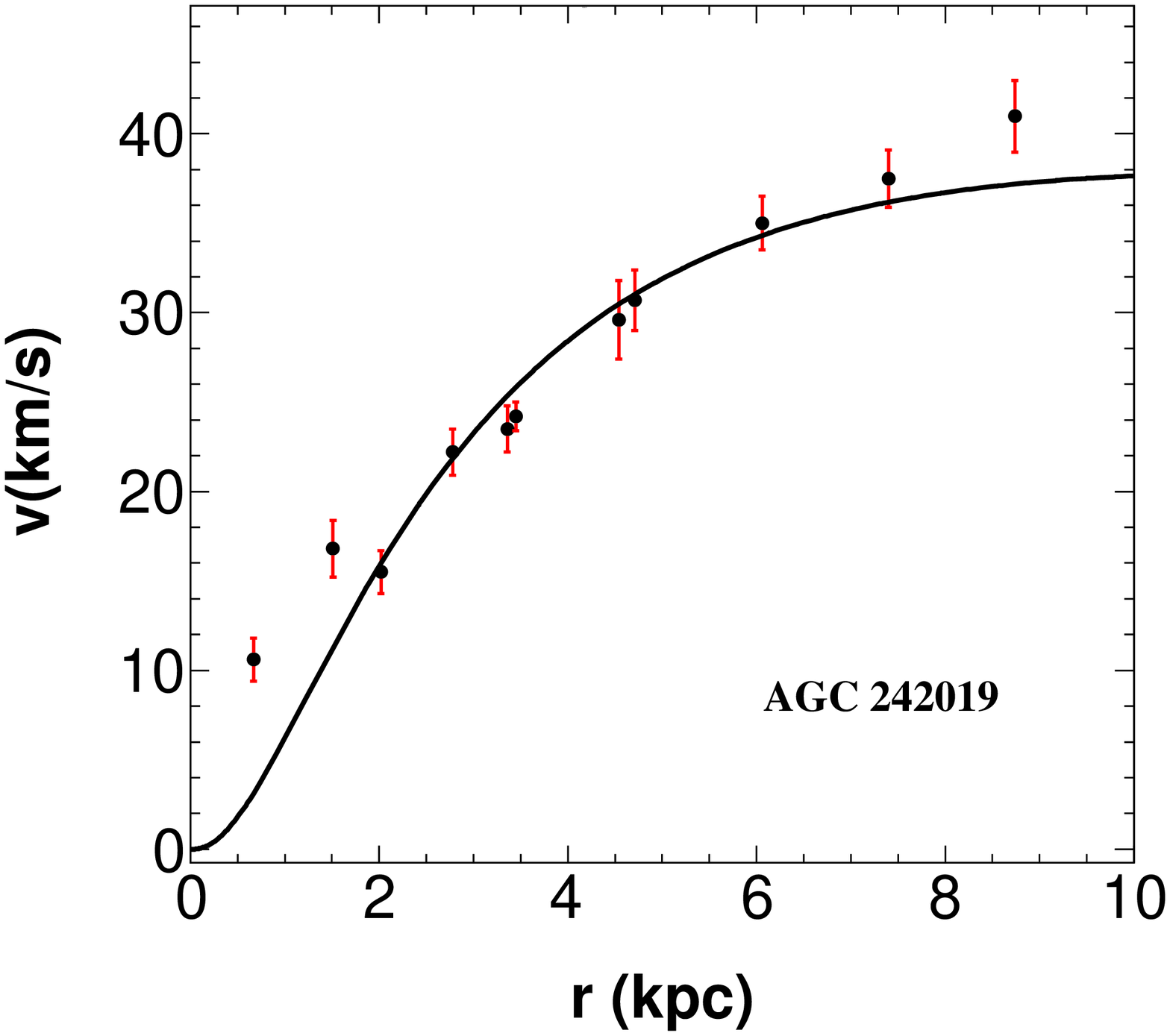}
    \caption{The rotational velocities of AGC $242019$ fitted to equation 
\eqref{eqn.30} with the associated errors plotted as a function of radial 
distance in kpc. The data for the rotational velocities of the galaxy is taken 
from Ref.~\cite{shi21}.}
    \label{fig.5}
\end{figure}

\section{The Tully-Fisher relation}
\label{sec.6}
R.~B.~Tully and J.~R.~Fisher in 1977 published a method of determining the 
distances of spiral galaxies based on their empirical or observational 
relation, now known as the Tully-Fisher relation \cite{tully}. It implies a 
relation between the luminosity of a galaxy and the velocity of the outermost 
observed point of the galaxy as
\begin{equation}
L = \xi\, v_{out}^a,
\label{eqn.32}
\end{equation}
where $L$ is the observed luminosity of a galaxy in units of 
$10^{10} L_\odot$, $v_{out}$ is the velocity at the outermost observed 
radial point of the galaxy in units of km/s, $\xi$ is the proportionality
constant and $a$ is another constant. Taking logarithm on both sides of this 
equation \eqref{eqn.32}, we may write
\begin{equation}
\log L = a \log v_{out} + b.
\label{eqn.33}
\end{equation}
Here $b = \log \xi$. We plotted the observed Tully-Fisher relation for the 
entire samples of galaxies used in this study as shown 
in the left panel of Fig.\ \ref{fig.5}. In the plot the vertical axis is the 
base $10$ logarithm of the $B$-band luminosity $L_B$ in units of $10^{10} 
L_\odot$ and the horizontal axis is the base $10$ logarithm of the velocity of 
the outermost observed radial point in units of kms$^{-1}$. The $B$-band 
luminosity data for each galaxy sample has been extracted from 
Ref.~\cite{mannheim} and are tabulated in Tables \ref{Table:1}, \ref{Table:2} 
and \ref{Table:3}. The solid line in the plot is the best least-square fitting 
of equation \eqref{eqn.33} to the observed corresponding data of galaxies as 
mentioned. The fitting is found to be very good with $\chi^2_{red} = 0.127$ 
and the parameters $a = 3.18\,\pm\, 0.27$ and $b = -\,6.75\,\pm\, 0.54$.        

\begin{figure}[!h]
\includegraphics[scale = 0.38]{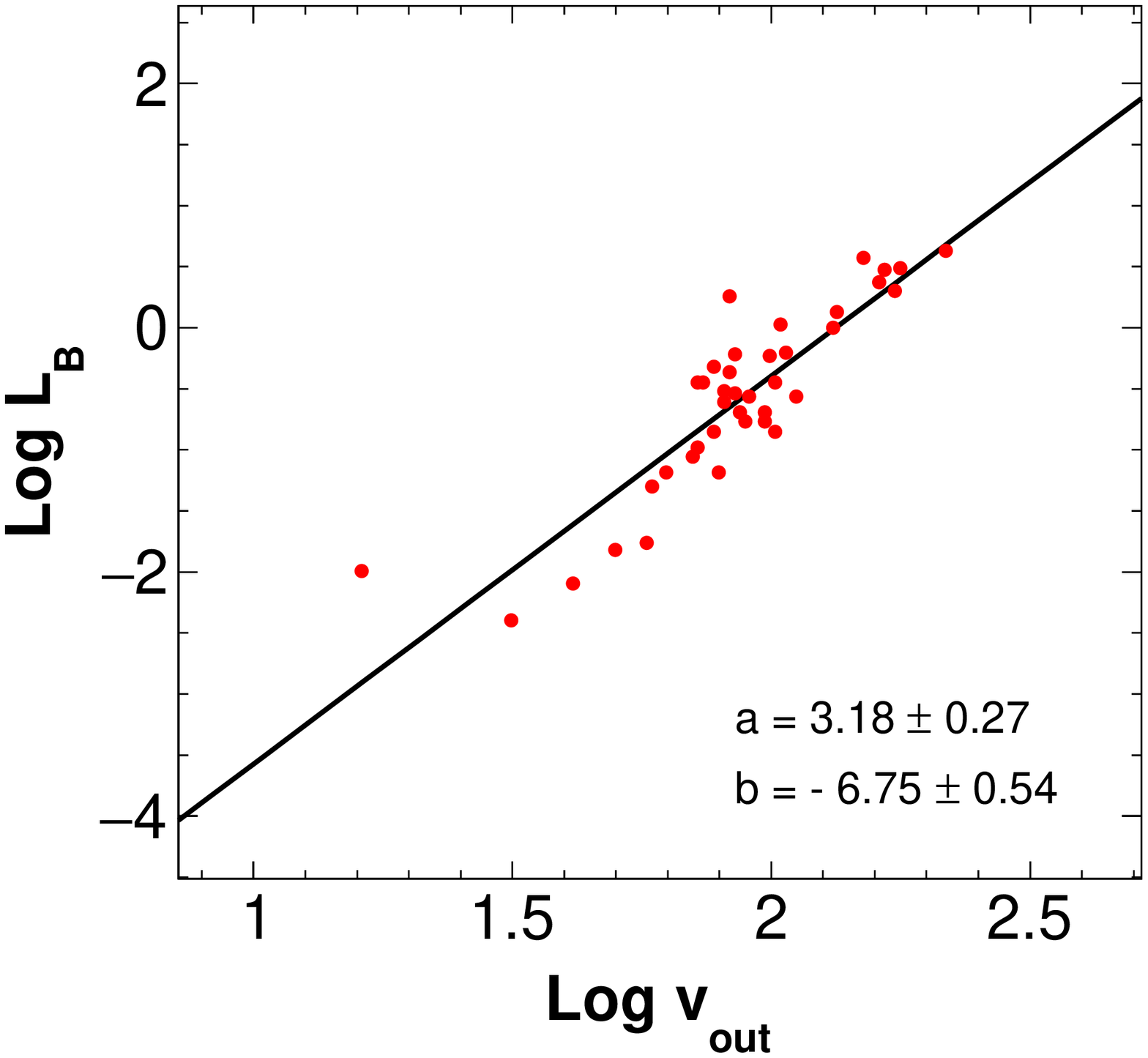} \hspace{0.2cm}
\includegraphics[scale = 0.38]{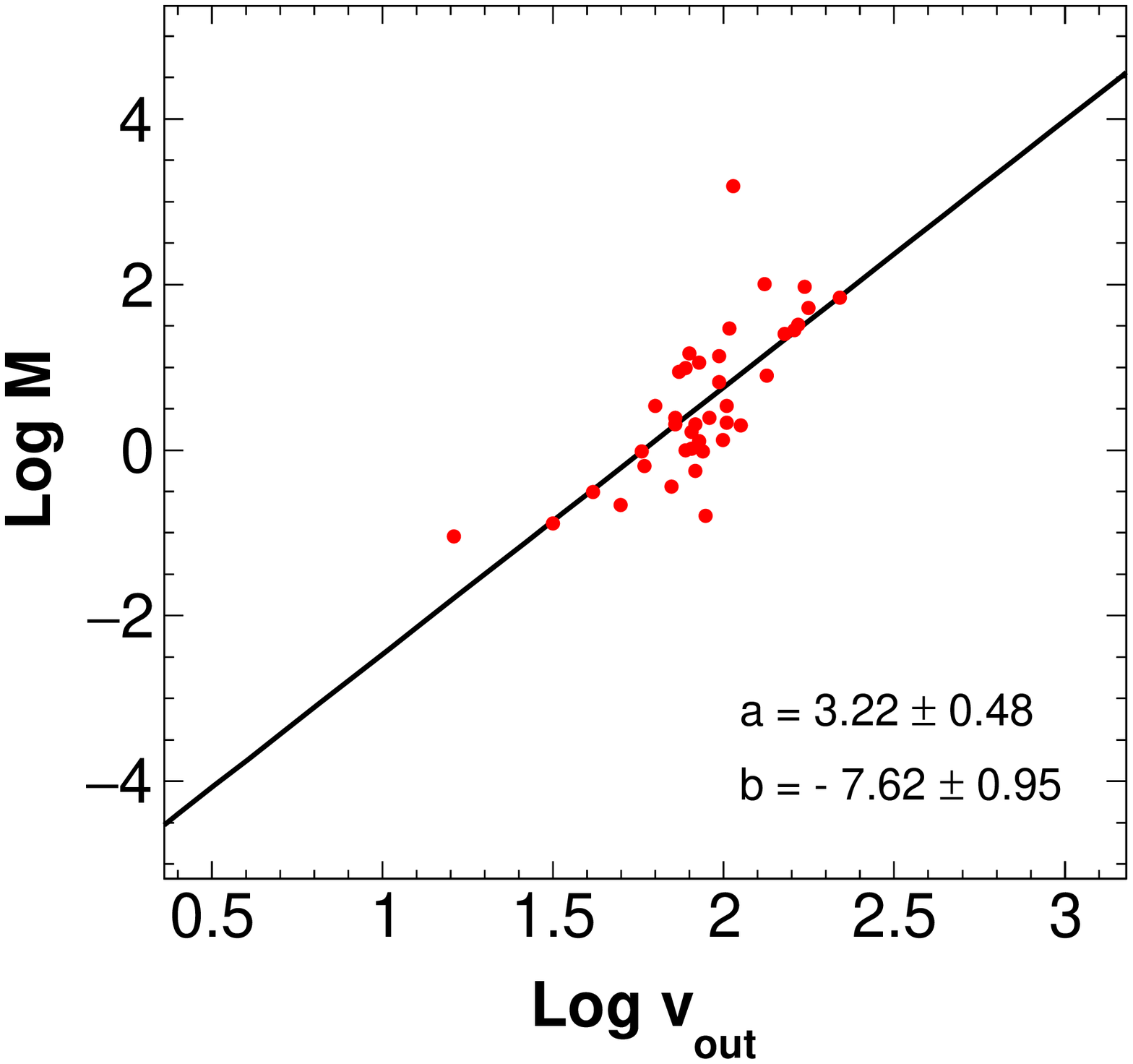} \hspace{0.2cm}
\caption{Plots for the Tully-Fisher relation. The left panel is for the 
observed $B$ band Tully-Fisher relation. The line in the plot is obtained by 
fitting equation \eqref{eqn.33} to the observed data. The right panel shows 
predicted Tully-Fisher relation for the masses and velocities of the outermost 
radial points of galaxies predicted from the new $f(\mathcal{R})$ 
model \eqref{eqn.27}. Here the solid line represents the best least-square fitting of
\eqref{eqn.34} to the corresponding predicted data as mentioned.}
\label{fig.6}
\end{figure}

Now, from the mean mass-luminosity relation $\langle M/L \rangle = constant$
\cite{brownstein} and equation \eqref{eqn.33} we can write the Tully-Fisher relation in 
terms of the predicted mass (in units of $10^{10} M_\odot$) and the velocity 
(in units of kms$^{-1}$ for the new $f(\mathcal{R})$ gravity model \eqref{eqn.27} as
\begin{equation}
\log M = a \log v + b + \log \langle M/L \rangle.
\label{eqn.34}
\end{equation}
The right panel of Fig.\ \ref{fig.5} shows the best fit Tully-Fisher relation 
parametrized by equation \eqref{eqn.34} for the new model \eqref{eqn.27}. The 
vertical axis denotes the base $10$ logarithm of the predicted mass (in units 
of $10^{10} M_\odot$) and the horizontal axis denotes the base $10$ logarithm 
of the fitted velocity (in units of kms$^{-1}$). In this fit the parameters 
are found to be $a = 3.22\,\pm\, 0.48$ and $b = -\,7.62\,\pm\, 0.95$ with 
$\chi^2_{red} = 0.397$.

\section{Summary and final remarks}
\label{sec.7}

The behaviour of galactic rotation curves indicates the need for DM or some 
modifications of GR. However, till date there is no direct evidence of the 
existence of DM. Also, DM interacts only via gravity and hence, the question 
arises as to whether the effect of DM is a consequence of the modification of 
gravity only. Keeping this point in mind, in this work, we have employed one 
of the simplest MTG, the $f(\mathcal{R})$ gravity to study the galactic 
rotation curves for some of the galaxies observed experimentally in recent 
times. In doing so we have mainly tried to test the viability of the recently 
proposed model \cite{dhruba, Dhruba} of $f(\mathcal{R})$ gravity 
in the galactic scales.

In our present work, we have used the Palatini formalism along with the Weyl 
transformation. Thus the variation of the action for the $f(\mathcal{R})$ 
theory first with respect to the metric and then with respect to the connection 
yields the field equations in the theory. We consider the Weyl transformation 
which is the frame transformation of the spacetime metric $g_{\mu\nu}$ from the 
Jordan frame to the Einstein frame to get the field equation in the convenient 
form of GR. Moreover, the connection in this torsion-free Palatini approach is 
the Levi-Civita connection for the conformally related metric and as such 
particles moving in the gravitational field follow the geodesics obtained by 
the connection. In the Einstein frame, we study the rotational velocity of a 
test particle moving in a stable circular orbit. A static spherically 
symmetric metric is taken into consideration and we are able to derive the 
co-efficients of the metric for our galactic model spacetime.

Next, we introduce the new model of $f(\mathcal{R})$ gravity. As 
mentioned already this is a recently introduced $f(\mathcal{R})$ gravity model 
and hence we try to test the viability of this model in the galactic scales. 
Making use of the metric co-efficients and the field equations we derive the 
analytical expression for rotational velocities of test particles in the 
stable circular orbits. It can be observed that the expression of the 
rotational velocity obtained is different from the Newtonian one. This is due 
to the fact that the rotational velocity obtained via our present approach 
has extra terms coming from the geometrical modifications in the theory. This 
rotational velocity expression is fitted with observed data of a few samples 
of galaxies. We consider nine samples of high surface brightness (HSB) 
galaxies, 21 samples of low surface brightness (LSB) galaxies and nine dwarf 
galaxies. The galactic rotation curves are well fitted with observations for 
the new $f(\mathcal{R})$ gravity model. Although, for a few samples of galaxies the 
mass-to-light ratios are found to be much larger than the expected upper 
bound, however the $\chi_{red}^2$ values for most of the samples are smaller 
or equivalent to 1, thus indicating the well fitted rotation curves. 
This shows the viability of the new $f(\mathcal{R})$ gravity model 
in the galactic scales.

In addition to these samples, we take into consideration an interesting class 
of galaxies i.e., the ultra diffuse galaxies (UDGs). They are fascinating 
objects which are either made entirely of DM or have a high content of DM. 
This makes them difficult to observe and analyze. We consider one such galaxy, 
AGC $242019$, which has been claimed to be similar to low surface brightness 
galaxies with a slowly rising rotation curve. Our study supports this 
proclamation. However, further studies are necessary in this regard. 

Finally, we have studied the Tully-Fisher relation in the new $f(\mathcal{R})$ gravity model. 
We have studied the luminosity as well as total mass of galaxies as a function 
of velocity of the outermost observed radial point. The entire sample of 
galaxies have been combined for this purpose and the fits show a consistent 
result across the galaxies. In our work, we have made use of the SPARC 
catalogue\ 
(\href{http://astroweb.cwru.edu/SPARC/}{http://astroweb.cwru.edu/SPARC/}) for 
the observational data of the sample of galaxies.

Lastly, it would be interesting to extend our work by including different
MTGs with more observational data and also by comparing our 
results with the standard DM profiles such as Navarro-Frenk-White (NFW) 
\cite{nfw, navarro} and Burkert \cite{burkert} profiles. Also, in the near 
future, we can try to obtain much feasible mass-to-light ratios within the 
expected bound.

\section*{Acknowledgments}
UDG is thankful to the Inter-University Centre for Astronomy and Astrophysics
(IUCAA), Pune, India for the Visiting Associateship of the institute.

\end{document}